\documentclass[12pt,halfline,a4paper]{ouparticle}

\usepackage{graphics} 
\usepackage{graphicx}
\usepackage{amsmath} 
\usepackage{amssymb}  
\usepackage[utf8]{inputenc}
\usepackage[english]{babel}
\usepackage[natbibapa]{apacite}
\usepackage{appendix}
\usepackage{arydshln}
\usepackage[labelfont=bf]{caption}
\usepackage{multirow}
\usepackage{float}
\usepackage{subfigure}
\usepackage{booktabs}

\begin{document}
	
\title{\textbf{\break {Stochastic Modeling and Forecast of Hydrological Contributions in the Colombian Electric System}}}
	
\author{%
	\name{Juan Pablo Pérez Monsalve}
	\address{Universidad EAFIT, Department of Finance\\
		Medellín, Colombia}
	\email{jperezm9@eafit.edu.co}
	\and
	\name{Freddy H. Marín Sanchez}
	\address{Universidad EAFIT, Department of Mathematical Sciences\\
		Medellín, Colombia}
	\email{fmarinsa@eafit.edu.co}}
	
%%%%%%%%%%%%%%%%%%%%%%%%%%%%%%%%%%%%%%%%%%%%%%%%%%%%%%%%%%%%%%%%%%%%%%%%%%%%%%%%

\abstract
	{In this paper as show that hydrological contributions in the colombian electrical system during the period between 2004 and 2016 have a periodic dynamic, with fundamental periods that are repeated every three years and that tend to oscillate around a long term average, in this context, such contributions can be characterized by mean reversion stochastic processes with periodic functional tendency. The objective of this paper is modeling and forecast the dynamic behavioral of hydrological contributions in the colombian electric system. A description of climate and hydrology in Colombia is carried out, as well as an analysis of periodicity and distributional properties of the data. A Gaussian estimation is performed, which allows to find all the constant and functional parameters from the historical data with daily frequency. The forecasts of the hydrological contributions are graphically illustrated for a period of three years and two alternatives to construct forecast fringes are proposed.} 
	
\date{}
	
	
\maketitle
	
%%%%%%%%%%%%%%%%%%%%%%%%%%%%%%%%%%%%%%%%%%%%%%%%%%%%%%%%%%%%%%%%%%%%%%%%%%%%%%%%
	
\section{Introduction}

South America has a great variety of geographical conditions and characteristics, represented by the different landscapes, which, together with the latitudinal extension, provides a high climatic diversity on the continent \citep{Jiminez2005}. Within this climatic diversity, the tropical zone emphasizes, where the great amount of solar energy avoids the existence of strong winters \citep{Latrubesse2005}, and its continental properties prevent the climatic events present in other places of the continent.

Colombia is a country where much of its territory has a tropical climate, specifically tropical humid, where rainfall usually exceeds 2000 mm for year. Its climate and weather are affected by factors such as the southern oscillation of the inter-tropical convergence zone (ITCZ), the Pacific and Atlantic oceans, the Amazon basin and the Andes mountain range, among others \citep{Marin2006}. All this climatic, geographical and biological configuration make Colombia one of the regions with the highest water resources in South America.

In order to take advantage of this surface water supply, the colombian electrical system is based on hydroelectric generation in about 80$\%$, where most of the generation stations are located in the Andean, Caribbean and Pacific region. The fact that the Colombian electrical system is focused on hydraulic generation, implies that the presence of changes that affect the hydric resource, causes variations of the level of water in the reservoirs, which translates into fluctuations in the price of electric energy. Therefore, the modeling and forecast of variables of climatic or hydrological order occupies an important place in the decision making of the agents of the electrical system.

The main investigations that address the forecast of climatic or hydrological variables of influence for the colombian electric sector are limited. In this sense, \cite{Salazar1998} use the singular spectral analysis, as well as the method of maximum entropy to adjust models of monthly precipitation for some basins in Antioquia, \cite{Poveda2002} evaluates different non-linear prediction methods of monthly average flows of 6 important rivers for the generation of electric energy in Colombia, \cite{Rojo-Hernandez2010} on the other hand, shows the nonlinear dynamics of the flows of the rivers of Colombia using a periodic model of prediction based on the singular spectral analysis. Thus, such investigations has been focused on specific hydrological regions, while using data that do not have high frequency, avoiding the capture of relevant information in the hydrological series like periodicity.

Particularly, the climatic and hydrological variability of Subtropical-Tropical South American region shows certain processes represented by low frequency quasi-periodical fluctuations \citep{Vargas2002}. This situation is not foreign to the contributions of water discharge of the rivers that contribute water to the reservoirs of the colombian electrical system, since, in daily frequency it is evidenced the existence of a fundamental period that is repeated every 3 years, in which there are sub-periods that are repeated every year.

The objective of this research is to model and forecast the water discharge of the rivers linked to the electric power generation system in Colombia, which exhibit periodic dynamics in their first logarithmic transformation. In this sense, we use the methodology proposed by \cite{Monsalve2017}, in which we estimate the parameters of one-factor mean stochastic reversion process, where the functional trend follows periodic behavior. For it, the estimation of maximum likelihood is used as well as the Fourier analysis. Subsequently, a period of the process is chosen, namely 2007-2010, with the information of this period we simulate paths for the period 2010-2013 and contrast them with the real data. 

This paper is organized as follows. In section \ref{climate} we explain the main climatic and hydrological conditions of Colombia, as well as a detailed description of the data used. In section \ref{textural} we present a textural modeling of the hydrological contributions in the colombian electric system, where we analyze its periodicity and its statistical and distributional properties. In section \ref{estimation} we find the parameters using a Gaussian estimation technique that which allows to perform a forecast of medium term. Finally section \ref{conclusions} shows the conclusions, comments and future works.  

\section{Climate and Hydrology in Colombia}\label{climate}

\subsection{Climate in Colombia}

The climate refers to atmospheric conditions prevailing in a place during a certain period \citep{Pabon-Caicedo2001}, whose characteristics are defined mainly by the astronomical location of that place, besides aspects such as general circulation, surface characteristics, altitude and exposure, among others \citep{Koppen1930}.

Astronomically, South America extends through approximately $65^{\circ}$ latitude (12 $^{\circ}$N-55 $^{\circ}$S) with a width extending along $45^{\circ}$ longitude (35 $^{\circ}$W-80 $^{\circ}$W). The tropical condition dominates most of the terrestrial surface and due to the small distance between the hottest and coldest months in the tropical parts and the attenuated range in the nontropical parts, there is no significant occurrence of the continental type climate \citep{Eidt1969}, and therefore the oceans are of great importance in most of the region.

In this sense, the wind coming from semi-permanent anticyclones linked to the Atlantic and Pacific ocean determines the air circulation. Thus, during the months of january and july these semi-permanent anticyclones change of position, being these seasonal changes fundamental determinants of the synoptic climatology of the region, which in turn take place due to the Intertropical Convergence Zone (ITCZ) where opposite effects of the trade winds of the northern hemisphere and the southern hemisphere are canceled.

The climate in the tropical region of South America is also affected by the geographic conditions of the region or surface characteristics, which cause changes in the dynamics of the wind circulation, where the most important is the Andean mountain chain. The Andes, prevent that the winds of the anticyclone of the Pacific South enter in the region, additionally, due to this mountain chain, humid winds from the east can transport heavy rain into the Amazon basin \citep{Eidt1969}. It is noteworthy that, due to the geographic conditions of the region, two factors, mountain-valley breeze and land-sea breeze, contribute to local climatic variations, which are caused by faster warming in the day and faster cooling of the air at night in the highest places.

The pattern of the different tropical climates occupies an important place in the formation of the hydrological cycle of such regions \citep{Balek1983}. Thus, the most variable element of the tropical climate corresponds to the rainfall, where three types, conventional, cyclonic and orographic are identified \citep{McGregor1998}, from these, the stations are determined, in such a way that, the quantity and the temporal distribution of such rainfalls are important criteria for distinguishing wet and dry sub-climatic zones \citep{Latrubesse2005}.

According to \cite{Ideam2005}, Colombia is a country where most of the region is classified as a per-humid and very humid climate according to the Thornthwaite climatic classification, the main areas included in this climate correspond to the Pacific, Orinoquia, Amazonia, and sectors of the Andean region in most of Antioquia, Caldas, Risaralda and western Santander. In turn, in the foothills of the three mountain ranges and to the south of the Caribbean region the climates are slightly humid, moderately humid and humid; while in the Cundiboyacense plateau, and sectors of the valleys of the high Magdalena and high Cauca, basins of the rivers Chicamocha and Zulia and sectors of the center of the Caribbean region, subhumid climates dry and subhumid humid are the constant.

The humid tropical climates are characterized by temperatures ranging from 24 to 30 $^\circ$C with an annual oscillation of around 3 $^\circ$C \citep{Latrubesse2005}. This situation is not foreign to the Colombian case, since in humid regions of Orinoquia, Pacific, and the Caribbean there are average temperatures near 28 $^\circ$C. In the Andean region temperatures vary according to height above sea level, slope exposure and precipitation regime, thus, for an elevation of 500 meters temperatures are between 25.2 $^\circ$C and 27.2 $^\circ$C for the eastern mountain range eastern slope and the western mountain range eastern slope respectively \citep{Ideam2_2005}. With respect to the circulation of the wind, Colombia is dominated by the trade winds characterized by its stability and by be weaks in general \citep{Ideam3_2005}. The highest winds are located in the region of Alta Guajira with average speeds of 6 meters for second, followed by regions such as the Caribbean coast, the north and south of the department of Cesar, among others, while in most of the country speed average annual wind is close to 2 meters for second. 

The factors described above make Colombia one of the countries with the greatest rainfall in the world, where also, there are inter-annual scale processes such as ENSO (El Niño oscillation of the south) events that influence the country's rainfall regime. Thus, \textit{El Niño} phenomenon causes a decrease in rainfall and an increase in temperatures, while \textit{La Niña} phenomenon has the opposite effect. The northern and central areas of the Pacific region have the highest rainfall in Colombia with an annual average between 8000 and 10000 mm, followed by the Amazon region with uniform rainfall throughout the year and average precipitations between 3000 and 4500 mm \citep{Ideam4_2005}. The rains in the Andean region are influenced by aspects such as geography and elevation, thus, the slopes located in the middle of the Magdalena and the middle Cauca, areas of the Eje Cafetero, Antioquia and Santander present the highest rainfalls with levels between 2000 and 4000 mm. It is precisely this geographic condition of the central and western   mountain ranges in the Andes that allows the presence of high volumes of rainfall in the extreme south of the Caribbean region (1800 to 2000 mm). 

\subsection{Hydrology in Colombia}

The climatic and geographical configuration of Colombia make it one of the countries with greater water wealth of the world. This richness is represented by the extensive surface water network, favorable groundwater storage conditions and the existence of large areas of wetlands, located in the greater part of the surface \citep{Cabrera2010}.

\begin{figure}[h]
	\centering
	\includegraphics[width=120mm]{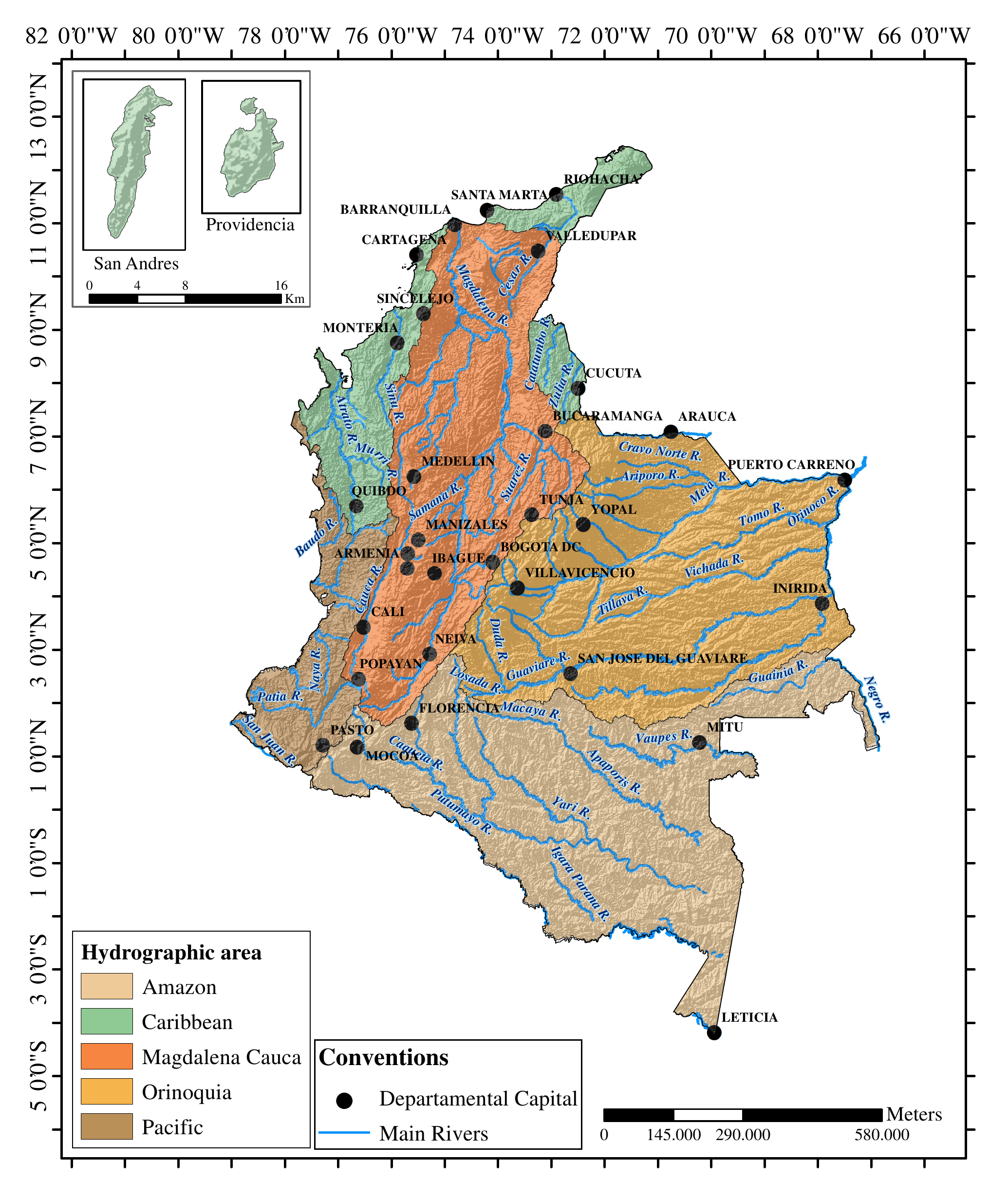}
	\caption{\textbf{Hydrographic Areas in Colombia}}
	\medskip
	\begin{minipage}{0.8\textwidth} 
		{\footnotesize{Hydrographic areas of Colombia according to the zoning and codification of hydrographic and hydrogeological units of Colombia made by the Instituto de Hidrología Meteorología y Estudios Ambientales (IDEAM).\par}}
	\end{minipage}
	\label{fig:1}
\end{figure}

The colombian hydrological regime is characterized by dry and humid periods during the year, where the water supply varies according to the five hydrographic areas of the country, the Caribbean, Magdalena-Cauca, Orinoquia, Amazon and Pacific, as shown in Figure \ref{fig:1}. Each one of these areas is conformed by hydrographic zones according to the composition of each region.

In this sense, the hydrology of the Caribbean region characterized by heterogeneity in its relief, is influenced by the Sierra Nevada of Santa Marta, the basins of the river Catatumbo, Rancheria, León and basin high and low of the Atrato river. In her, the highest hydrological contributions come from the Sinu river in the department of Cordoba and the Atrato river in the departments of Chocó and Antioquia. Thus, the percentage of water supply in this area represents 9.1$\%$ of the national supply during a year, while its water discharge is 5,799 $m^{3}/s$ \citep{Santos2014}.

In the Magdalena-Cauca hydrographic area, there is an important supply of surface waters in the upper, middle and lower basins of the Magdalena river and the Cauca river basin. Thus, rivers such as Suaza, Páez, Cabrera, Saldaña, Coello and Bogotá contribute with important volumes of water to the upper basin of the Magdalena River, while the low basin is fed by rivers such as Gualí, Cimitarra, Lebrija, Chicamocha, Sogamoso, Suárez, among others. In the lower Magdalena, on the other hand, the rivers Cauca and San Jorge converge to the river Magdalena. The Cauca river basin is characterized by a variety of hydro-climatic systems, where the Tarazá, Nechí and Porce rivers provide high volumes of water. The hydric supply in this region equals 13.5$\%$ at the national level, with a water discharge of 8,595 $m^{3}/s$ \citep{Santos2014}.

The Orinoquia region is represented by the high basins of the Arauca and Casanare rivers, where most of its rivers originate in the foothills of the eastern cordillera. In this region, large rivers flow, as the Arauca, Meta, Guaviare, that by their length and volume become navigable during most of the year \citep{Garcia2010}. The hydric supply is 26.3$\%$ in relation to the national percentage, while its water discharge is 529,469 $m^{3}/s$ \citep{Santos2014}.

The hydrographic area of the Amazon is composed of the basins of the rivers Amazonas, Caquetá, Putumayo, Vaupés and Guainía whose rivers are mighty. This region exhibits an extensive tropical forest and a variety of ecosystems that are configured to create a high biodiversity. The hydric supply is the highest in Colombia representing 37$\%$ of the total, as well as the water discharge with 23,626 $m^{3}/s$ \citep{Santos2014}.

For its part, the Pacific region exhibits the highest rainfall and water yields in the country, which is mainly made up of the basins of the rivers Patía, San Juan, Micay, Baudó and Atrato. The previous conditions make the hydric supply in this region equivalent to 14.1$\%$ of the national total, with a water discharge of 8,980 $m^{3}/s$ \citep{Santos2014}.

\subsection{Data}

The objective of this research is to model and forecast the hydrological contributions in the colombian electrical system. For this purpose, we take as reference variable, the water discharge of the rivers linked to the Sistema Interconectado Nacional (SIN), which in turn is the integrative model that allows the operation of the electric sector in Colombia from generation to transmission and distribution of electricity. Thus, the variable water discharge measured in $m^{3}/s$ is obtained from the BI portal of the company XM which is the company that operates and manages the Colombian electricity market. 

\begin{figure}[htbp]
	\centering
	\subfigure[Hydrological Regions]{\includegraphics[width=50mm]{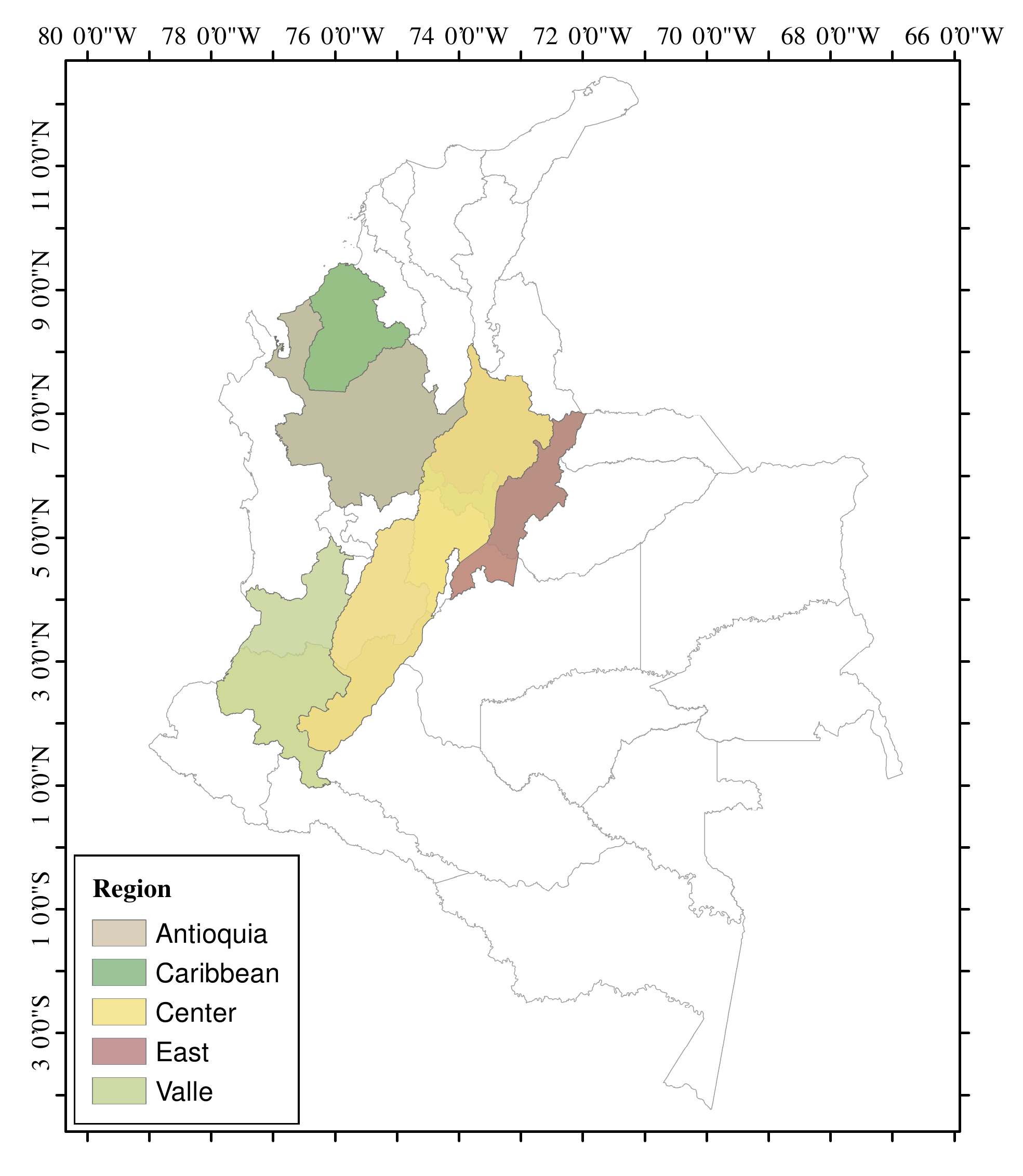} \label{fig:2-1}}
	\subfigure[Antioquia]{\includegraphics[width=78mm]{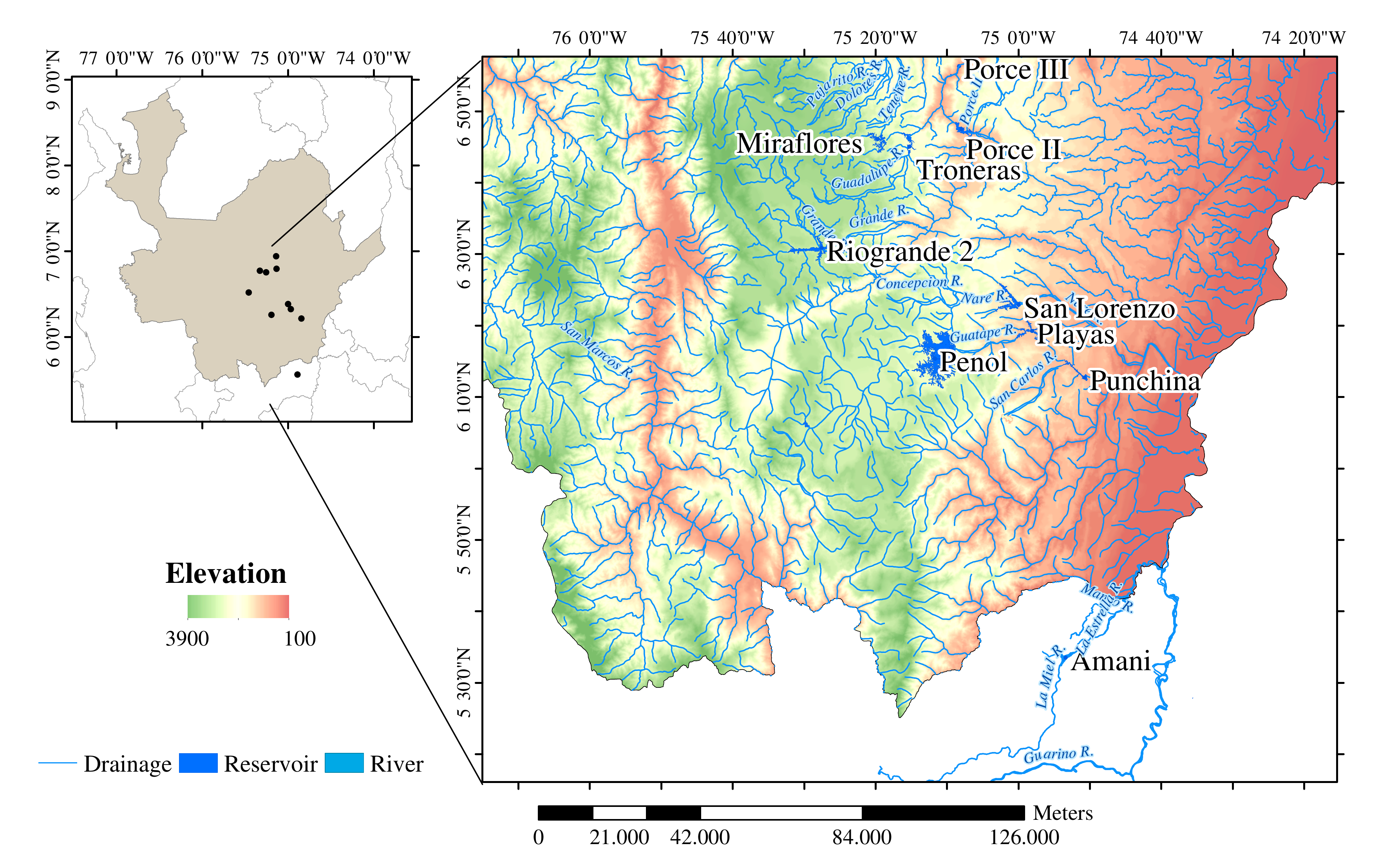}\label{fig:2-2}}	
	\subfigure[Caribbean]{\includegraphics[width=78mm]{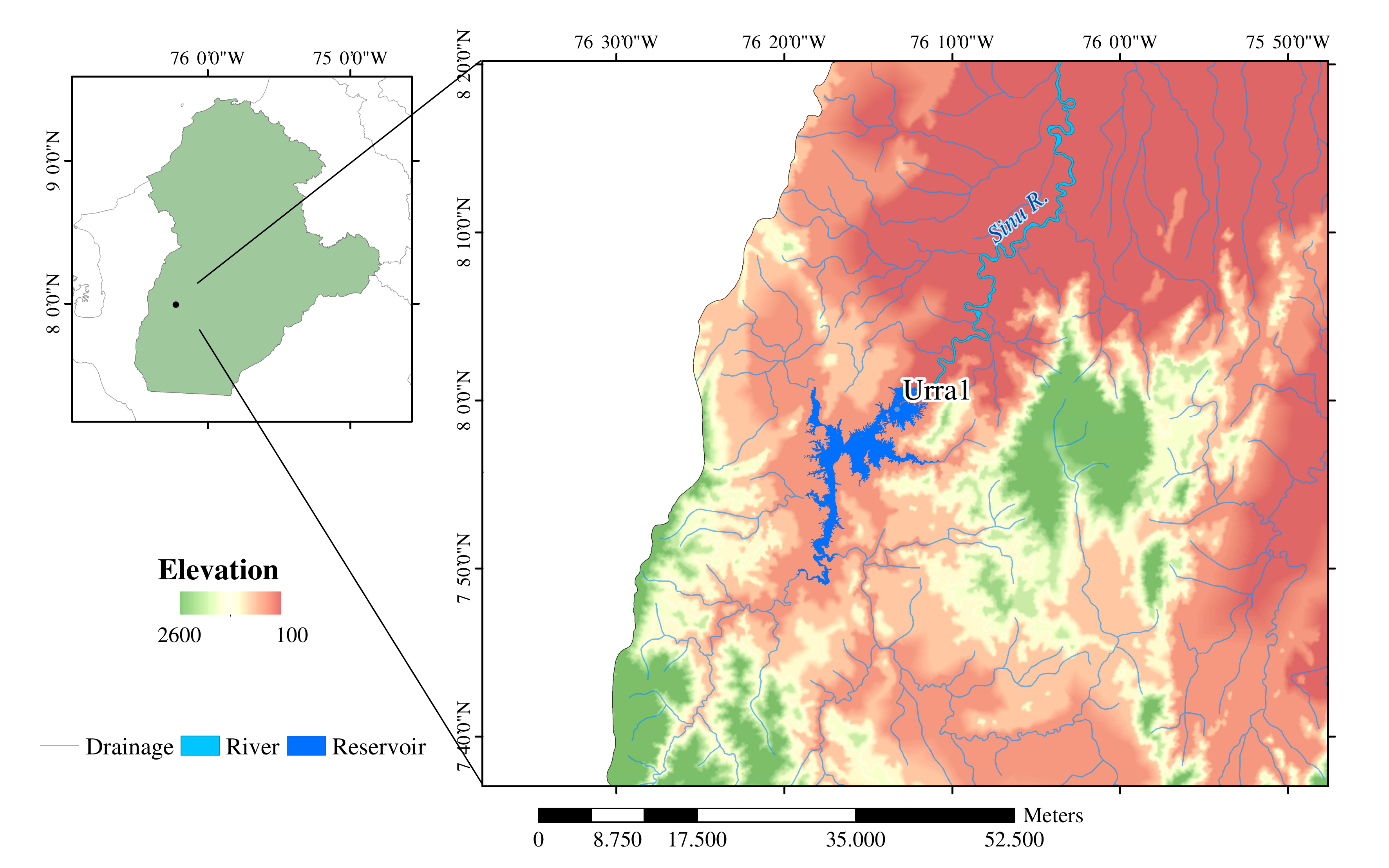}\label{fig:2-3}}
	\subfigure[Center]{\includegraphics[width=78mm]{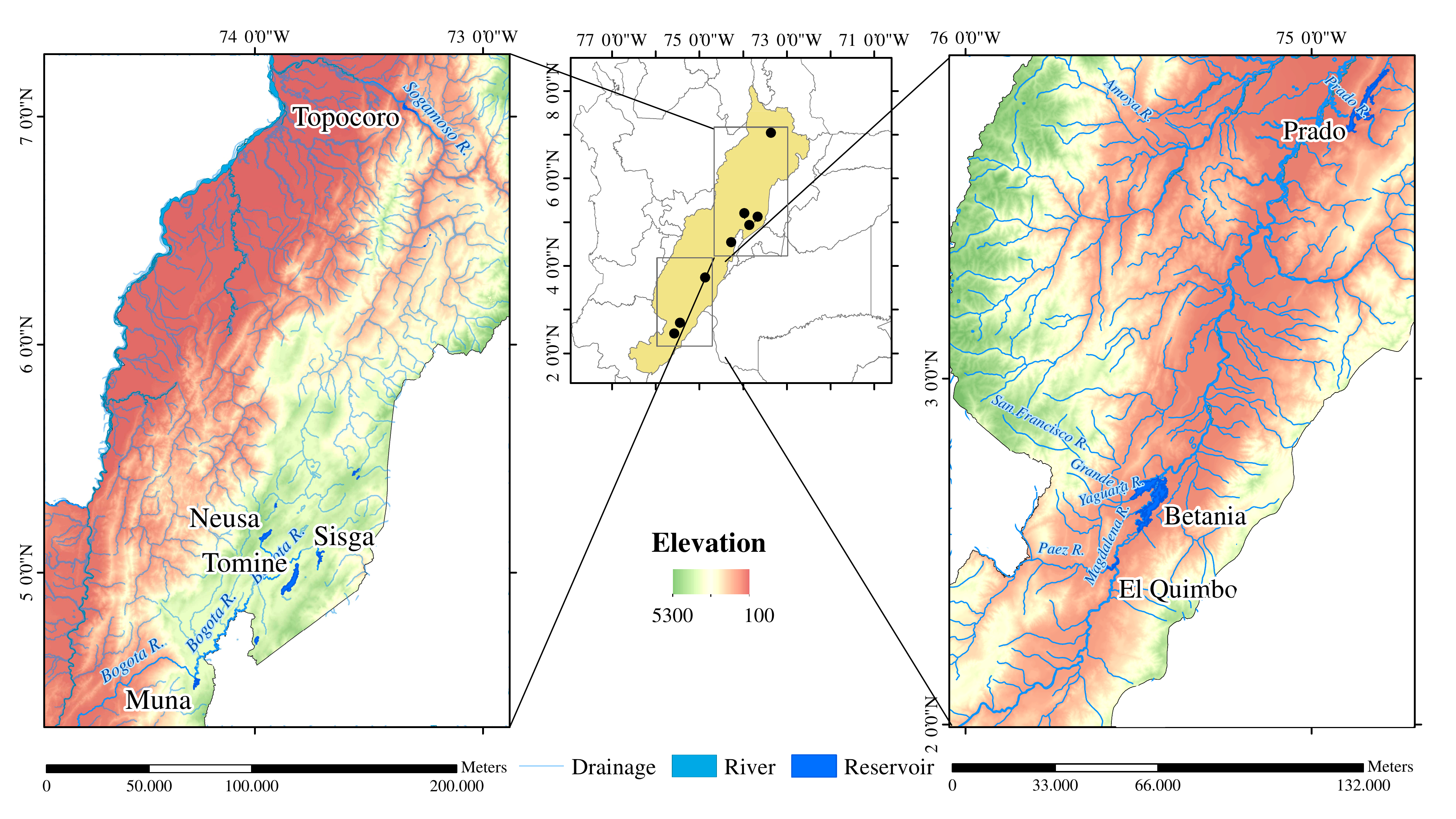}\label{fig:2-4}}
	\subfigure[East]{\includegraphics[width=78mm]{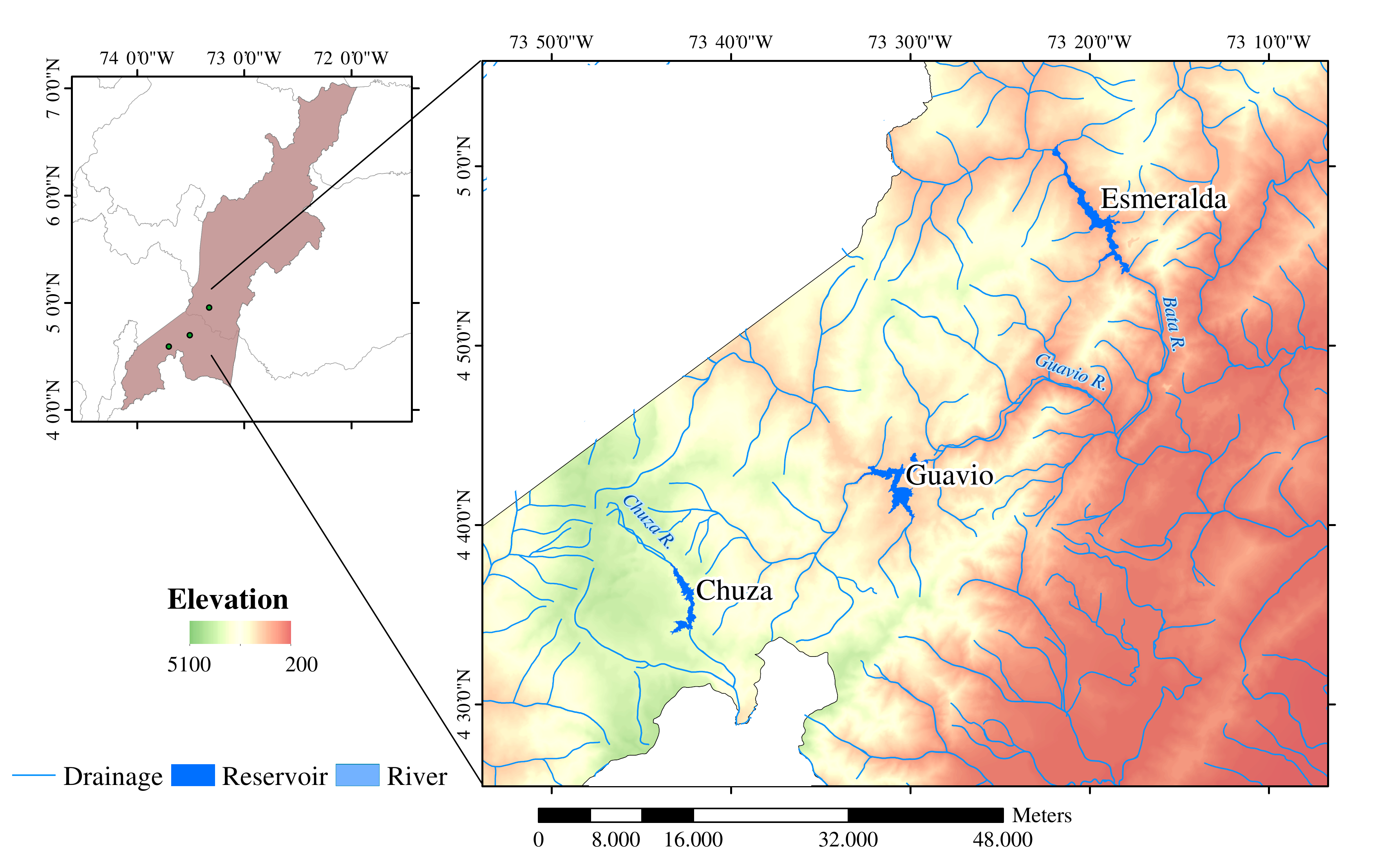}\label{fig:2-5}}
	\subfigure[Valle]{\includegraphics[width=78mm]{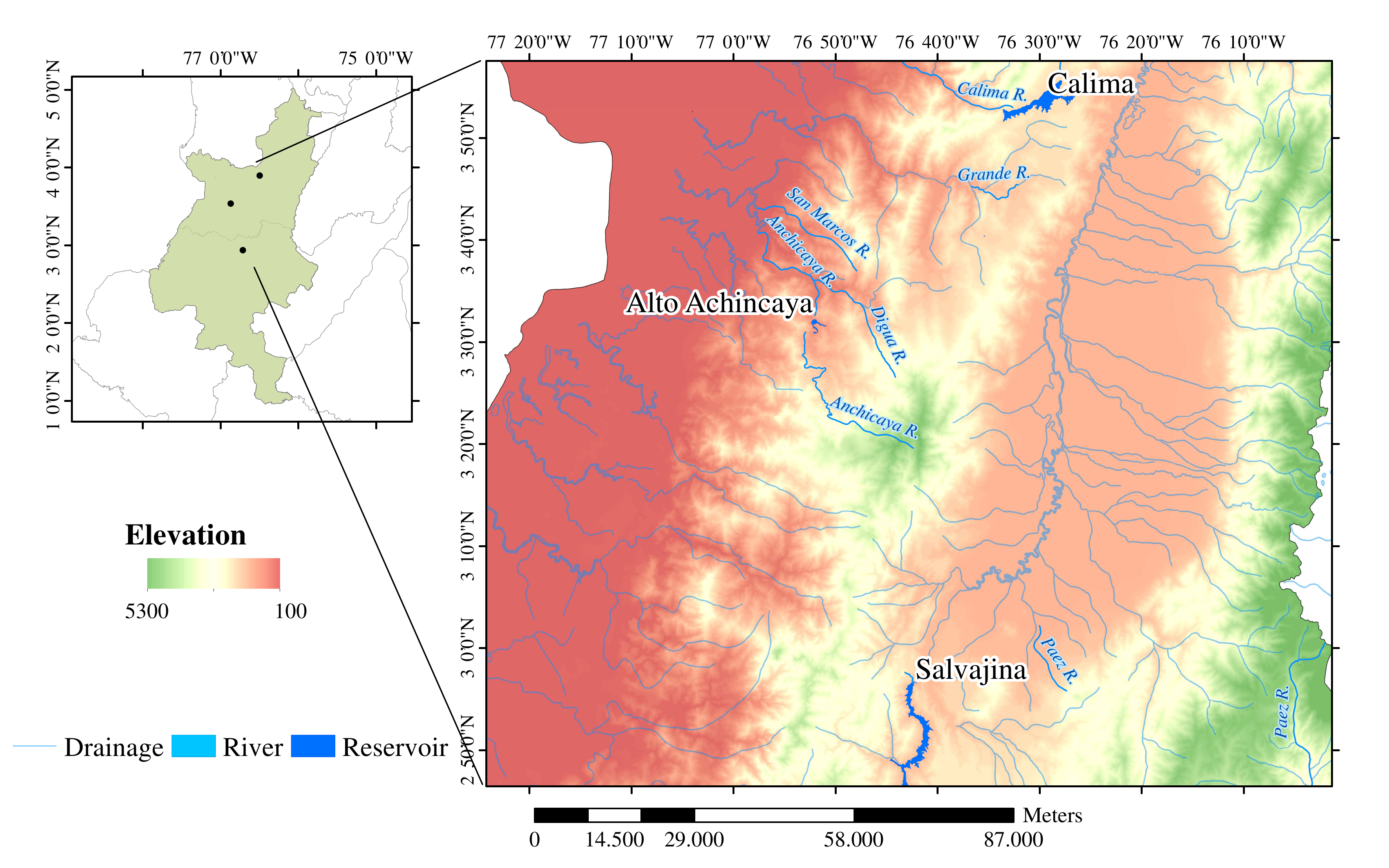}\label{fig:2-6}}
	\caption{\textbf{Hydrological Regions in the Colombian SIN}} \label{fig:2}
	\medskip
	\begin{minipage}{0.8\textwidth} 
		{\footnotesize{Hydrological regions in Colombia linked to SIN. The reservoirs and rivers of the 5 regions, Antioquia, Caribbean, Center, East and Valle are observed.\par}}
	\end{minipage}
\end{figure}

XM therefore receives, collects and manages the information of the rivers that contribute water to the reservoirs of the Colombian electrical system, information that is provided by the companies that own or operate these reservoirs and that intervene in the generation of hydroelectric energy. In this sense, in Colombia, there are about 25 reservoirs that are part of the electrical system, which are located in various places of the national geography and are nourished by different rivers. According to this location five hydrological regions are defined, namely, Antioquia, Caribbean, Center, East and Valle as shown in Figure \ref{fig:2}.

Antioquia, defined in Figure \ref{fig:2-2}, is one of the regions with the highest hydrographic representation in the colombian electrical system with important reservoirs such as Peñol, Playas, Miraflores, among others. The Caribbean region represented in Figure \ref{fig:2-3}, thanks to its geographic, climatic and environmental characteristics, has only the Urrá 1 reservoir, whose water resource comes mainly from the Sinú river. On the other hand, in the Central region (Figure \ref{fig:2-4}) rivers such as Betania, Prado and Quimbo flow, among others, allowing the formation of reservoirs bearing the same name. The East hydrological region as shown in Figure \ref{fig:2-5} is formed by three reservoirs, Esmeralda, Chuza and Guavio, while the Valle region (Figure \ref{fig:2-6}) has the influx of the Calima, Cauca Salvajina, Alto Achincaya and Digua rivers and the Calima, Salvajina and Alto Achincaya reservoirs.

Table \ref{tab:hidros} specifies in greater detail the water resources in the Colombian SIN for each hydrological region, as well as the methodology and frequency of water discharge measurement of the rivers considered. Most operators or owners of the reservoirs use the water balance or operating balance to measure the water discharge of the rivers that supply to these reservoirs, while some others use direct measurement depending on the reservoir. The water balance is based on the principle of mass conservation, where what enters in the reservoir (water discharge, rainfall, water imports, etc.) less what comes out of the reservoir (for power generation, dumping, water exports, etc.) equals the storage difference in said reservoir, this difference in volume (expressed in cubic meters) is adjusted to the time and other variables to obtain the water discharge. The direct method consists in the dam site measurement, thus, with a station of measurement of the level of the river in the tail of the reservoir, the level of entry and the level transited to the point of dam are calculated, using regression techniques or by area factor.

\begin{table}[htbp]
	\centering
	\caption{\textbf{Hydrological Sources in the Colombian SIN}}
	\resizebox{\textwidth}{!}{
		\begin{tabular}{ccccccccc}
			\toprule
			\toprule
			\textbf{Region} &       & \textbf{Reservoir} &       & \textbf{River} &       & \textbf{Meas. Methodology} &       & \textbf{Meas. Frequency} \\
			\cmidrule{1-1}\cmidrule{3-3}\cmidrule{5-5}\cmidrule{7-7}\cmidrule{9-9}    \multirow{14}[1]{*}{\textbf{Antioquia }} &       & San Lorenzo &       & A. San Lorenzo &       & Direct measurement  &       & Horary \\
			&       & Troneras  &       & Concepción &       & Direct measurement  &       & Daily \\
			&       & Troneras  &       & Deviations EEPPM &       & Direct measurement  &       & Daily \\
			&       & Riogrande 2 &       & Grande &       & Water balance &       & Daily \\
			&       & Troneras  &       & Guadalupe &       & Water balance &       & Daily \\
			&       & Playas &       & Guatape &       & Water balance &       & Daily \\
			&       & Amani &       & Miel I &       & Direct measurement  &       & Horary \\
			&       & Amani &       & Deviations Guarino &       & Direct measurement  &       & Horary \\
			&       & Amani &       & Deviations Manso &       & Direct measurement  &       & Horary \\
			&       & Peñol &       & Nare  &       & Water balance &       & Daily \\
			&       & Miraflores &       & Tenche &       & Water balance &       & Daily \\
			&       & Punchina &       & San Carlos &       & Direct measurement  &       & Horary \\
			&       & Porce II &       & Porce II &       & Direct measurement  &       & Daily \\
			&       & Porce III &       & Porce III &       & Water balance &       & Daily \\
			\hdashline
			&       &       &       &       &       &       &       &  \\
			\textbf{Caribbean} &       & Urra 1 &       & Sinú  &       & Water balance &       & Horary \\
			\hdashline
			&       &       &       &       &       &       &       &  \\
			\multirow{9}[0]{*}{\textbf{Center}} &       & Tominé  &       & \multirow{4}[0]{*}{Bogotá N.R., Blanco } &       & \multirow{4}[0]{*}{Water balance} &       & \multirow{4}[0]{*}{Daily} \\
			&       & Neusa &       &       &       &       &       &  \\
			&       & Sisga  &       &       &       &       &       &  \\
			&       & Muña  &       &       &       &       &       &  \\
			&       & -     &       & Amoya &       & -     &       & - \\
			&       & Betania &       & Betania CP &       & Direct measurement  &       & Every 15 minutes \\
			&       & Prado &       & Prado &       & -     &       & - \\
			&       & Topocoro &       & Sogamoso &       & Direct measurement &       & Horary  \\
			&       & El Quimbo &       & El Quimbo  &       & Direct measurement &       & Every 15 minutes \\
			\hdashline
			&       &       &       &       &       &       &       &  \\
			\multirow{3}[0]{*}{\textbf{East}} &       & Esmeralda &       & Bata  &       & -     &       & - \\
			&       & Chuza &       & Chuza &       & Water balance &       & Daily \\
			&       & Guavio &       & Guavio &       & Water balance &       & Daily \\
			\hdashline
			&       &       &       &       &       &       &       &  \\
			\multirow{4}[1]{*}{\textbf{Valle}} &       & Calima  &       & Calima &       & Water balance &       & Horary \\
			&       & Salvajina &       & Cauca Salvajina &       & Water balance &       & Horary \\
			&       & Alto Achincaya &       & Alto Achincaya &       & Water balance &       & Horary \\
			&       & Alto Achincaya &       & Digua &       & Water balance &       & Horary \\
			\bottomrule
			\bottomrule
			\multicolumn{9}{p{20cm}}{\footnotesize{The main rivers and reservoirs of the SIN are presented, information obtained from the company XM. Also we shows the methodology of measurement and measuring frequency for the water discharge of rivers, information obtained from the companies that own or operate the reservoir of interest. EEPPM indicates Empresas Públicas de Medellín.}}
	\end{tabular}}
	\label{tab:hidros}
\end{table}

In this context, we take the water discharge information ($m^{3}/s$) that operators or owners of the reservoirs report daily to the company XM. Subsequently we make an aggregation of all the data, obtaining then the total water discharge for the SIN, to finally make a first logarithmic transformation on the data, which will be the reference variable in the subsequent sections and will be called hydrological contributions. The chosen time window is located between february 5, 2004 and february 5, 2016, giving a total of 6192 observations with daily frequency. The choice of this period is due to the possible existence of periodicity in the data.

\section{Stochastic Modeling}\label{textural}

\subsection{Periodicity}

Various natural, physical and financial phenomena are described by processes where periodicity appears implicitly. However, with the exception of experiments in controlled environments the fluctuations of such phenomena are hardly periodic, hence, the almost periodicity becomes important to describe more precisely the dynamics of these phenomena \citep{Bezandry2011,Diagana2007}.

\begin{figure}[h]
	\centering
	\includegraphics[width=\textwidth]{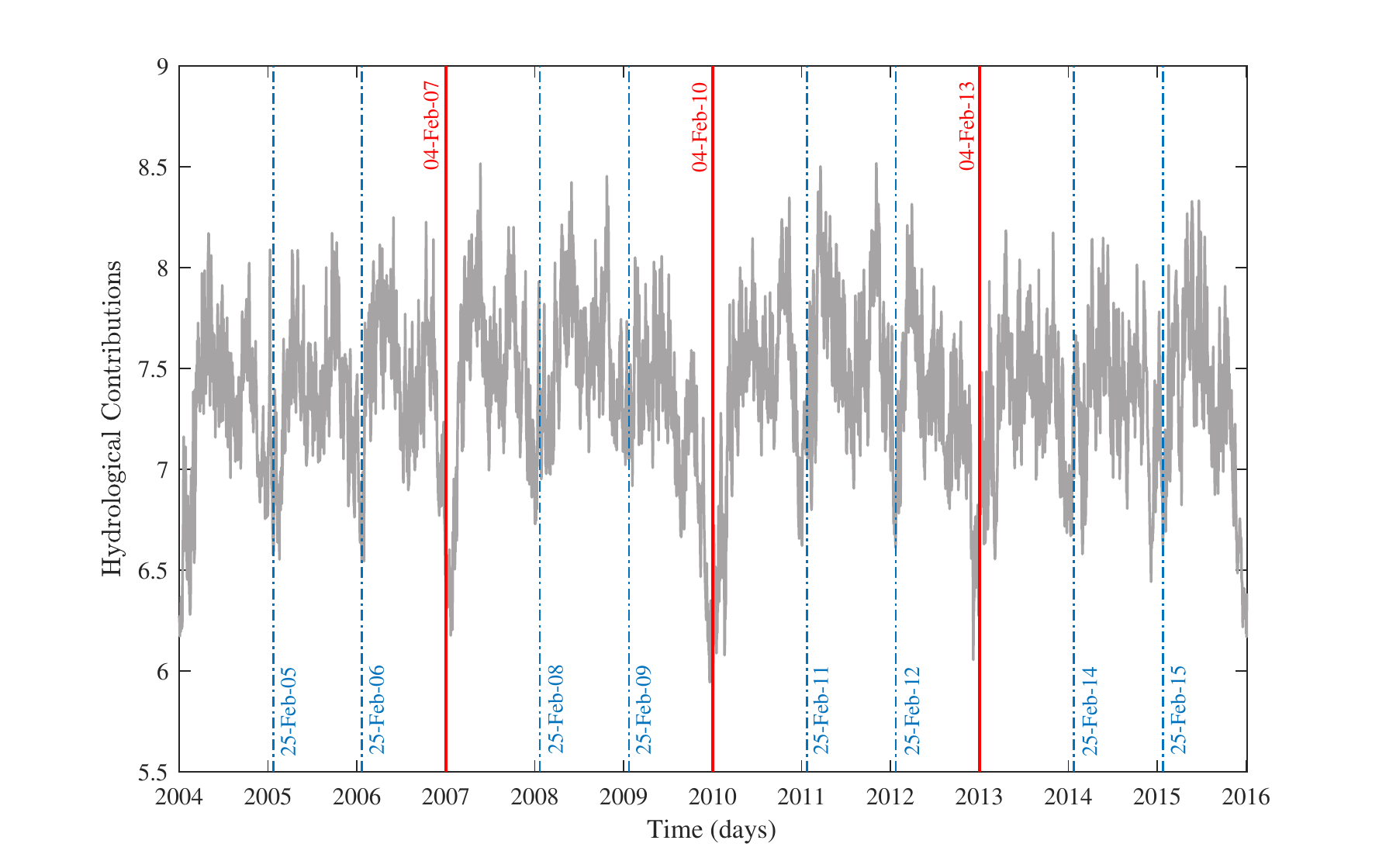}
	\caption{\textbf{Periodic Decomposition of Hydrological Contributions}}
	\medskip
	\begin{minipage}{0.8\textwidth} 
		{\footnotesize{Colombian daily hydrological contributions for the period from February 5, 2004 to February 5, 2016. The continuous red lines divide the periods every three years approximately, while the blue dotted sub-periods of one year approximately.\par}}
	\end{minipage}
	\label{fig:3}
\end{figure}

\begin{figure}[h]
	\centering
	\includegraphics[width=\textwidth]{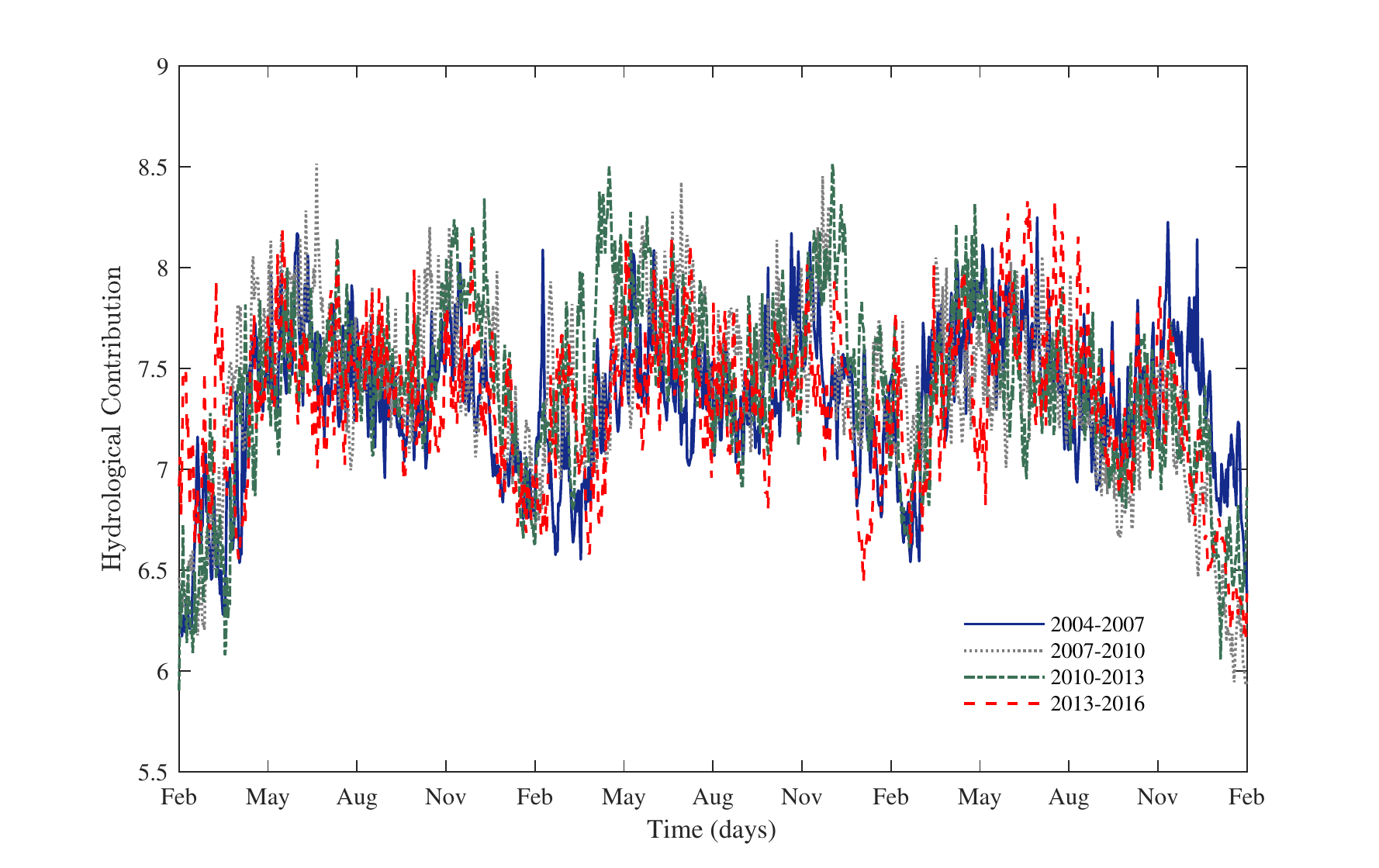}
	\caption{\textbf{Overlapping Periods for Hydrological Contributions}}
	\medskip
	\begin{minipage}{0.8\textwidth} 
		{\footnotesize{Colombian daily hydrological contributions for the period from February 5, 2004 to February 5, 2016, where each period is an interval of three years\par}}
	\end{minipage}
	\label{fig:4}
\end{figure}

In terms of climatology, most of the phenomena have a dynamic that presents a regular pattern and is repeated over a fixed period, such as temperature during periods of seasons, levels of precipitation, among others, situation that is not foreign to the Colombian case. In particular, the hydrological  contributions appear to exhibit periodic dynamics as shown in Figure \ref{fig:3}. Thus, the graphical analysis indicates a fundamental period that is presented every three years for a date close to February 4, as the continuous red lines denote it. In addition, each fundamental period presents sub-periods that are repeated approximately every year, as dotted blue lines denote it. This hypothesis is observed in greater detail overlapping each one of the fundamental periods, as shown in Figure \ref{fig:4}.

The periodogram in spectral analysis is a tool of recurrent use to identify periodicities, this estimates the function of spectral density for a continuous set of frequencies, estimation based on the spectral representation theorem where the autocovariance function and the spectral density function are transformed of Fourier one of the other \citep{Madisetti1998}. In the case of hydrological contributions, there are no pronounced peaks that indicate possible periods. However, it may be misleading to look for peaks in a periodogram, since, the ordinates at the Fourier frequencies are relatively independent, they are bound to fluctuate and show many small peaks and troughs \citep{Bloomfield2004}. For this reason, \cite{Fisher1929} proposed a test of significance of the largest peak in the periodogram through the g-statistic, expressed by \cite{Wichert2004} such as,

\begin{equation}
g=\frac{\max_{k}I(\omega_{k})}{\sum_{k=1}^{[N/2]}I(\omega)_{k}}
\end{equation}

Where, $I(\omega)_{k}$ denotes the periodogram and $N$ the sample size. Thus, large values of $g$ reject the null hypothesis according to which the process is purely random. For the case of the Colombian hydrological contributions, we assume the methodology proposed by \cite{Wichert2004}, with 4 time series that correspond to the fundamental periods, for each one, the Fisher's g-statistic and its respective p-value are calculated. The results shown in the Table \ref{tab:g-estadistico} indicate that the four selected periods reject the null hypothesis according to which they follow a purely random dynamics, therefore, such periods have a statistically significant periodic component.

\begin{table}[htbp]
	\centering
	\caption{\textbf{Fisher's Periodicity Test}}
	\begin{tabular}{ccccccccc}
		\toprule
		\toprule
		\textbf{Statistic} &       & \textbf{2004-2007 } &       & \textbf{2007-2010 } &       & \textbf{2010-2013 } &       & \textbf{2013-2016 } \\
		\cmidrule{1-1}\cmidrule{3-3}\cmidrule{5-5}\cmidrule{7-7}\cmidrule{9-9}   
		\textbf{g-statistic} &       & 0.806814 &       & 0.832377 &       & 0.771508 &       & 0.786318 \\
		\textbf{p-value} &       & 0.000000 &       & 0.000000 &       & 0.000000 &       & 0.000000 \\
		\bottomrule
		\bottomrule
		\multicolumn{9}{p{14.5cm}}{\footnotesize{g-statistic of Fisher and its p-value for the hydrological contributions in Colombia, calculated with the procedure proposed by \cite{Wichert2004} with $N=1036$ and $G=4$.}}
	\end{tabular}
	\label{tab:g-estadistico}
\end{table}

In this context, the hydrological contributions in Colombia exhibit a periodic dynamic, with a fundamental period present every 3 years, so the first is located approximately between 5-Feb-2004 and 4-Feb-2007, the second between 5- Feb-2007 and 4-Feb-2010, the third between 5-Feb-2010 and 4-Feb-2013, and the fourth between 5-Feb-2013 and 4-Feb-2016. In addition, within each fundamental period there are sub-periods that are repeated each year during a date close to February 25.

\subsection{Statistical and Distributional Properties}

Hydrological contributions in Colombia have exhibited an average behavior close to 7.4, where the period that has presented the lowest average level has been between 2013 and 2016, while the period between 2010 and 2013 has had the highest average level, as shown in Table \ref{tab:descrip}. In terms of dispersion from the first period (2004-2007) to the third (2007-2010) there is an increase in the variation starting at 0.36 and ending at 0.44. For its part, the Augmented Dickey-Fuller test indicates that the data considered are not stationary and therefore it is necessary to take a first difference. The first difference of hydrological contributions shows that all periods are stationary at around 0\%, where the standard deviation is at levels close to 0.18. 

\begin{table}[htbp]
	\centering
	\caption{\textbf{Descriptive Statistics}}
	\resizebox{\textwidth}{!}{
		\begin{tabular}{ccccccccccccc}
			\toprule
			\toprule
			\textbf{Variable} &       & \multicolumn{3}{c}{\textbf{Statistic}} &       & \textbf{2004-2007} &       & \textbf{2007-2010} &       & \textbf{2010-2013} &       & \textbf{2013-2016} \\
			\cmidrule{1-1}\cmidrule{3-5}\cmidrule{7-7}\cmidrule{9-9}\cmidrule{11-11}\cmidrule{13-13}    \multirow{3}[1]{*}{\textbf{HC}} &       & \multicolumn{3}{c}{\textbf{Mean}} &       & 7.3335 &       & 7.3864 &       & 7.4064 &       & 7.3244 \\
			&       & \multicolumn{3}{c}{\textbf{Standar D.}} &       & 0.3660 &       & 0.4323 &       & 0.4386 &       & 0.3809 \\
			&       & \multicolumn{3}{c}{\textbf{ADF Test}} &       & -0.3418 &       & -0.4799 &       & -0.5230 &       & -0.3899 \\
			&       &       &       &       &       &       &       &       &       &       &       &  \\
			\multirow{7}[1]{*}{\textbf{Difference}} &       & \multicolumn{3}{c}{\textbf{Mean}} &       & 0.0001 &       & -0.0005 &       & 0.0009 &       & -0.0005 \\
			&       & \multicolumn{3}{c}{\textbf{Standar D.}} &       & 0.1733 &       & 0.1703 &       & 0.1807 &       & 0.1805 \\
			&       & \multicolumn{3}{c}{\textbf{ADF Test}} &       & -36.7300*** &       & -38.7572*** &       & -37.1779*** &       & -35.2690*** \\
			&       & \multirow{2}[0]{*}{\textbf{Normality Period}} &       & \multicolumn{1}{l}{\textbf{Start}} &       & 23-Dec-2004 &       & 16-Nov-2007 &       & 03-Mar-2011 &       & 14-Jun-2014 \\
			&       &       &       & \multicolumn{1}{l}{\textbf{End}} &       & 14-Dec-2006 &       & 02-Mar-2009 &       & 04-Feb-2013 &       & 25-Jan-2016 \\
			&       & \multicolumn{3}{c}{\textbf{Jarque-Bera}} &       & 4.3818 &       & 4.3062 &       & 4.3173 &       & 2.8371 \\
			\bottomrule
			\bottomrule
			\multicolumn{13}{p{20.5cm}}{\footnotesize{HC denotes the hydrological contributions, Difference is the first difference of hydrological contributions. *** denotes 99\% confidence level. For the Jarque-Bera test the null hypothesis indicates normality, while for ADF (Augmented Dickey-Fuller) the null hypothesis is that there is a unit root.}}
	\end{tabular}}
	\label{tab:descrip}
\end{table}

\begin{figure}[h]
	\centering
	\subfigure[2004-2007]{\includegraphics[width=70mm]{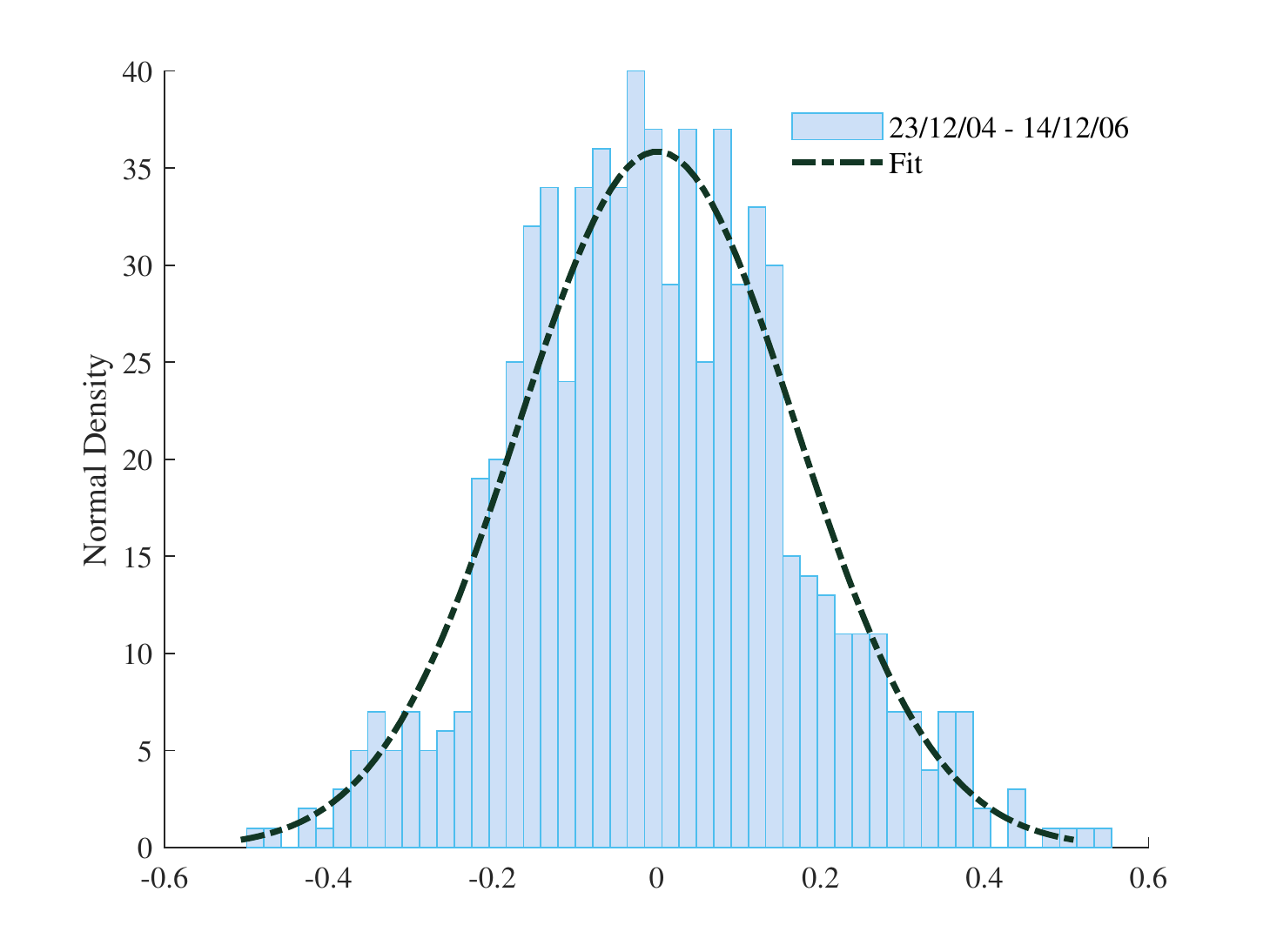}\label{fig:5-1}}
	\subfigure[2007-2010]{\includegraphics[width=70mm]{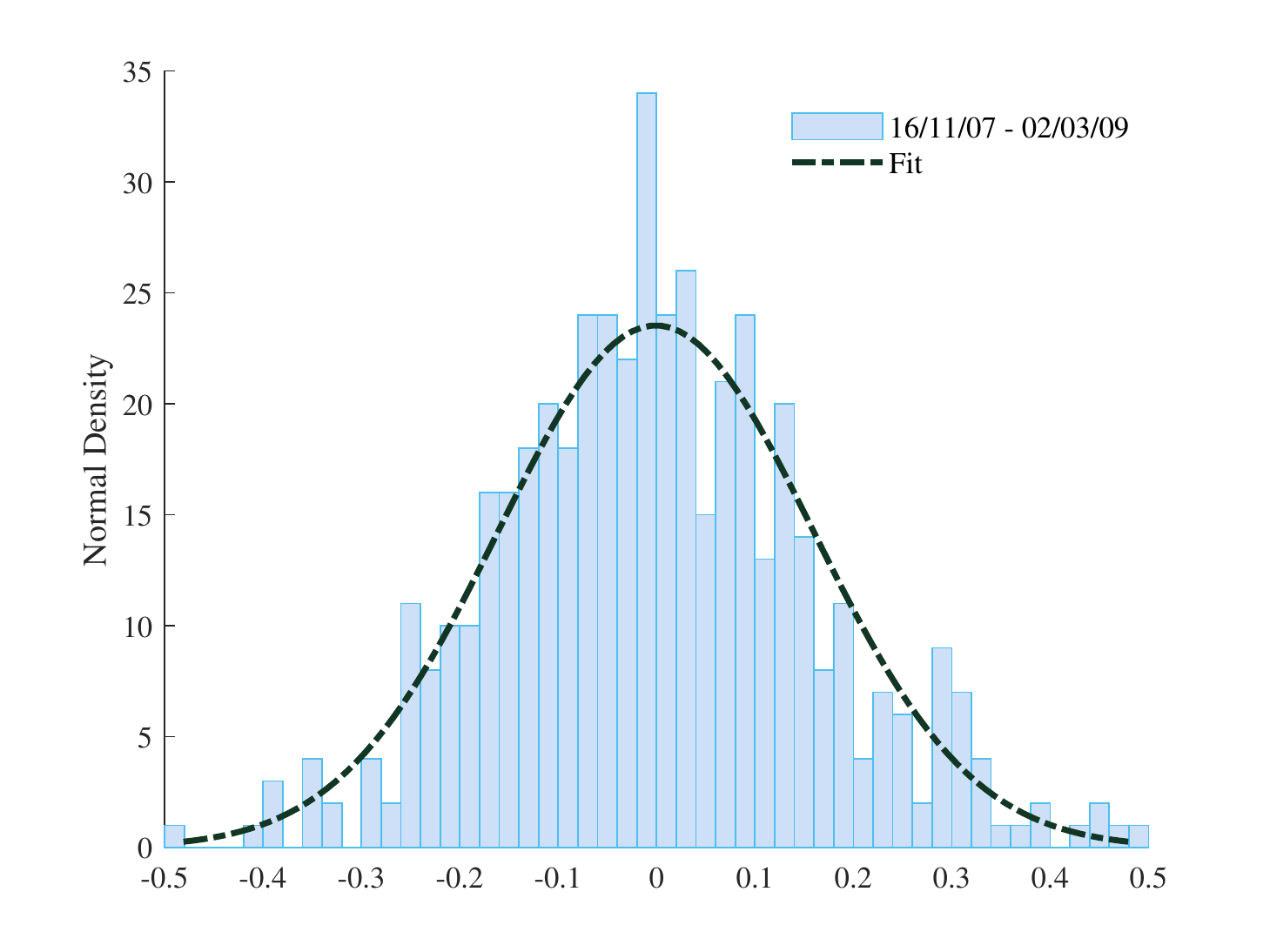}\label{fig:5-2}}
	\subfigure[2010-2013]{\includegraphics[width=70mm]{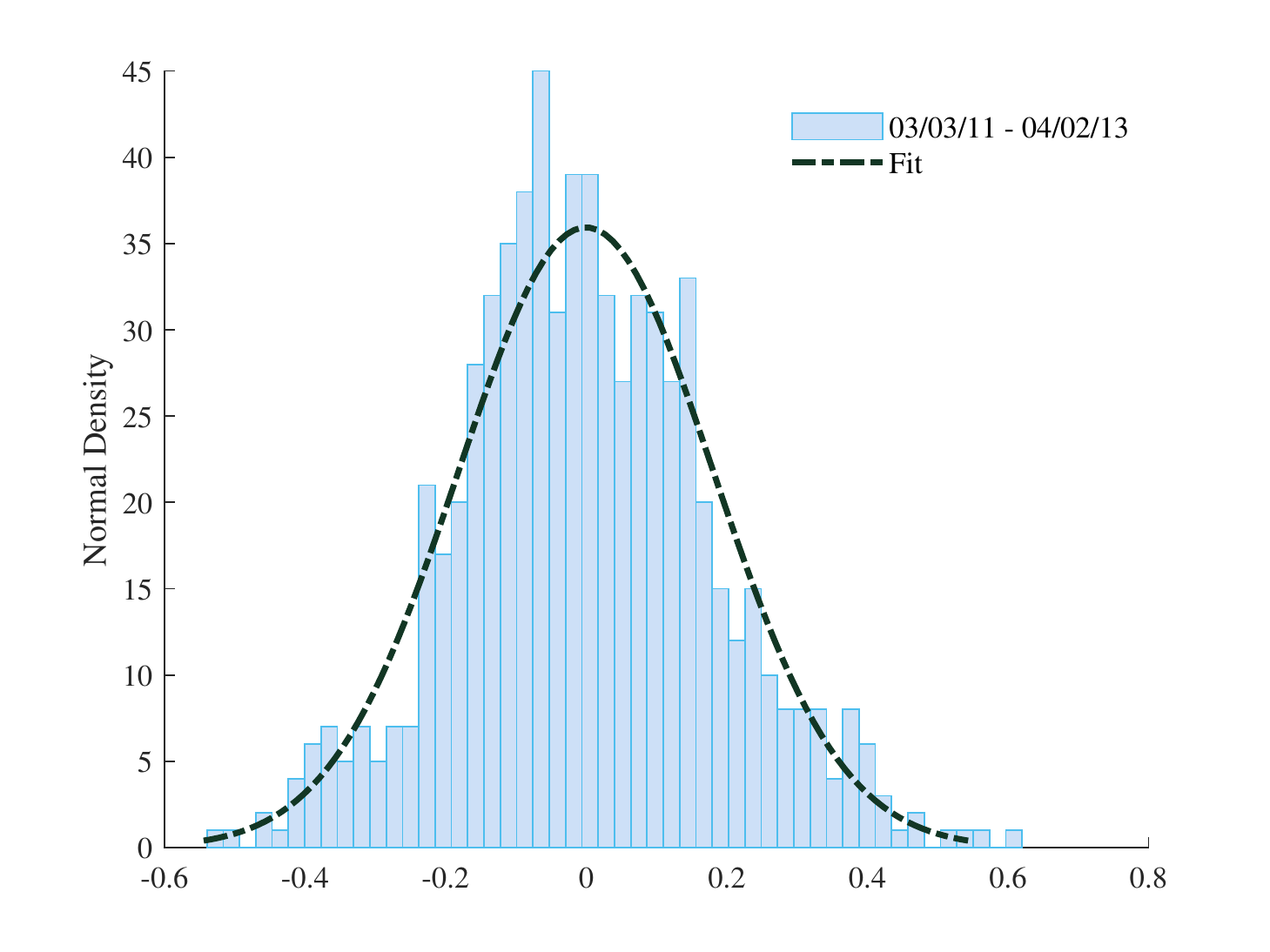}\label{fig:5-3}}
	\subfigure[2013-2016]{\includegraphics[width=70mm]{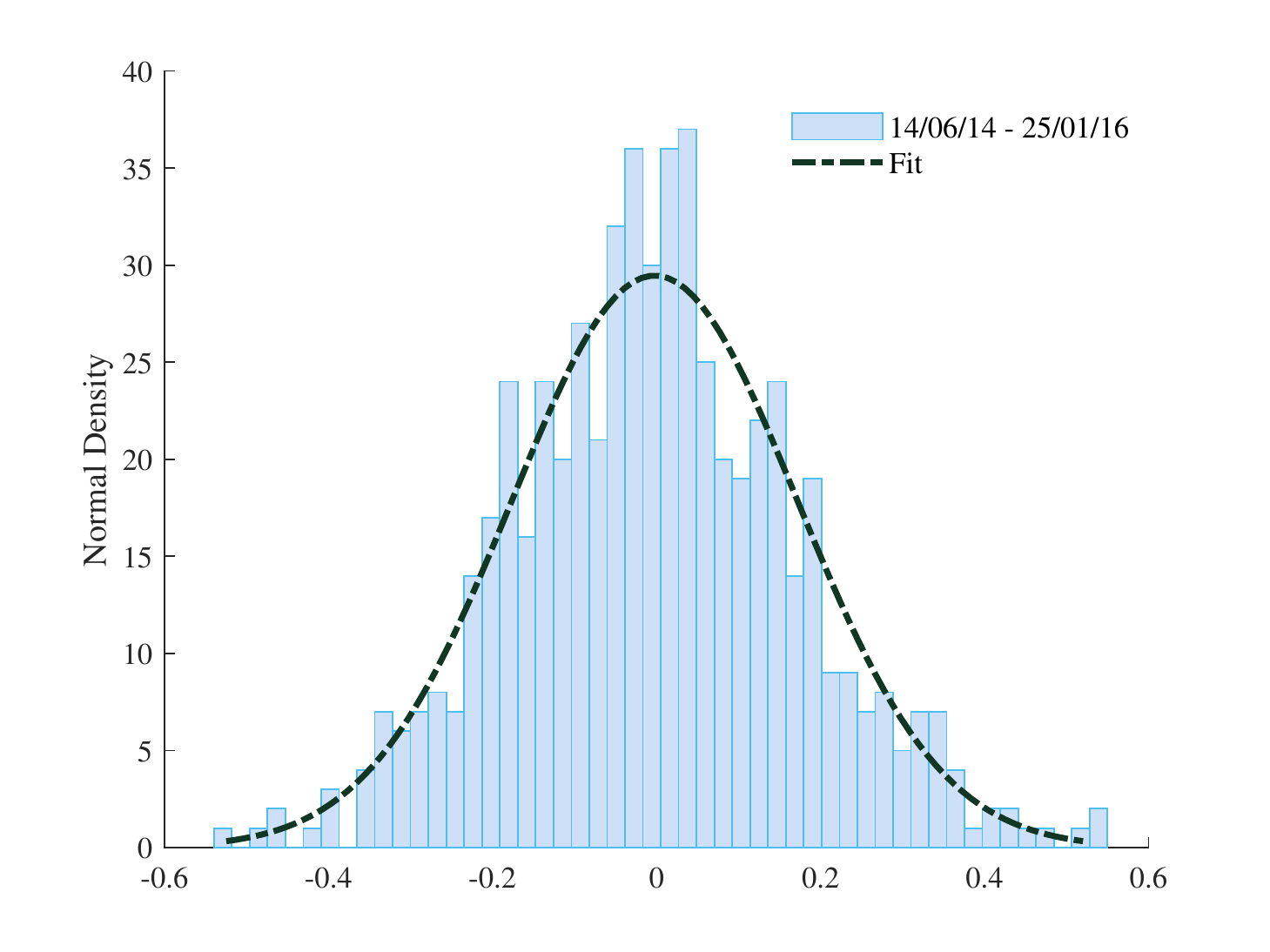}\label{fig:5-4}}
	\caption{\textbf{Normal Density of the Difference of Hydrological Contributions}} \label{fig:5}
	\medskip
	\begin{minipage}{0.8\textwidth} 
		{\footnotesize{Normal density for the intervals of each period defined in Table \ref{tab:descrip}. The histogram of the data and the normal theoretical distribution that corresponds to them are shown.\par}}
	\end{minipage}
\end{figure}

On the other hand, the analysis of distributional properties on the difference indicates that for each study period the normality hypothesis is satisfied. To guarantee this hypothesis, we selected test intervals within each fundamental period, in which increased or reduced the level of water in the reservoirs are presented. Therefore, the choice of these intervals was made considering scenarios of drought and abundant rains, that resulted in low and top peaks in the data. The results of the analysis on the selected intervals indicate that the normality hypothesis is accepted by the test of Jarque-Bera as shown in Table \ref{tab:descrip}. Figure \ref{fig:5} shows the histogram for the interval of each period and its normal theoretical distribution, which results confirm the mentioned previously. 

To verify if the hydrological contributions revert to the mean we use the Variance Ratio test (VR). In this test the null hypothesis indicates that the data follow a random walk such that the variance of the return of $k$-period is $k$ times the variance of the one-period return and, therefore, the expected value of VR($k$) must be equal to unity for all horizons $k$ \citep{Malliaropulos1999}. Thus, data will revert to the mean if VR($k$) is significantly lower than the unit at long horizons, $k$ \citep{Malliaropulos1999}. The results of the Variance Ratio for all periods as shown in Table \ref{tab:meanrevertion}, indicate that the hydrological contributions revert to the mean since for different horizons.

\begin{table}[h]
	\centering
	\caption{\textbf{Variance Ratio Test}}
	%\resizebox{\textwidth}{!}{
	\begin{tabular}{ccccccccccc}
		\toprule
		\toprule
		\multirow{2}[4]{*}{\textbf{Period}} &       & \multirow{2}[4]{*}{\textbf{Statistic}} &       & \multicolumn{7}{c}{$k$} \\
		\cmidrule{5-11}          &       &       &       & \textbf{2} &       & \textbf{4} &       & \textbf{8} &       & \textbf{16} \\
		\cmidrule{1-1}\cmidrule{3-3}\cmidrule{5-5}\cmidrule{7-7}\cmidrule{9-9}\cmidrule{11-11}          &       &       &       &       &       &       &       &       &       &  \\
		\multirow{2}[0]{*}{\textbf{2004-2007}} &       & \textbf{VR} &       & 0.8961*** &       & 0.6541*** &       & 0.4136*** &       & 0.2720*** \\
		&       & \textbf{z-Statistic} &       & -3.4387 &       & -6.1184 &       & -6.5597 &       & -5.4731 \\
		&       &       &       &       &       &       &       &       &       &  \\
		\multirow{2}[0]{*}{\textbf{2007-2010}} &       & \textbf{VR} &       & 0.8439*** &       & 0.6476*** &       & 0.4514*** &       & 0.2991*** \\
		&       & \textbf{z-Statistic} &       & -5.1663 &       & -6.2331 &       & -6.1376 &       & -5.2693 \\
		&       &       &       &       &       &       &       &       &       &  \\
		\multirow{2}[0]{*}{\textbf{2010-2013}} &       & \textbf{VR} &       & 0.8829*** &       & 0.6711*** &       & 0.4746*** &       & 0.2966*** \\
		&       & \textbf{z-Statistic} &       & -3.8869 &       & -5.7010 &       & -5.6760 &       & -5.0877 \\
		&       &       &       &       &       &       &       &       &       &  \\
		\multirow{2}[1]{*}{\textbf{2013-2016}} &       & \textbf{VR} &       & 0.9359*** &       & 0.7396*** &       & 0.5120*** &       & 0.3308*** \\
		&       & \textbf{z-Statistic} &       & -1.9284 &       & -4.2490 &       & -5.1156 &       & -4.7732 \\
		\bottomrule
		\bottomrule
		\multicolumn{11}{p{15.5cm}}{\footnotesize{Variance Ratio test (VR) and its z statistic for all fundamental periods of hydrological contributions. The calculations were made with heteroskedasticity robust standard error estimates, where the results are the same with homoskedasticity. *** denotes $99\%$ confidence level. The null hypothesis is that the data behave as a random walk.}}
	\end{tabular}
	\label{tab:meanrevertion}
\end{table}

\section{Estimation and Forecast}\label{estimation}

\subsection{Model and Method of Estimation}

In order to ensure that daily data on the hydrological contributions of the Colombian electrical system can be modeled as a mean reversion process with periodic functional tendency, as defined in \cite{Monsalve2017}, it is necessary to ensure periodicity, normality in the first difference, and reversion to the mean. As shown above, such contributions follow a periodic dynamic with a fundamental period of three years duration and sub-periods of annual frequency. With regard to normality, we find that in each period the normality hypothesis for the difference is satisfied through the Jarque-Bera test. On the other hand, the results of the Variance Ratio test indicate that the variable considered presents reversion to the mean for all the fundamental periods. Thus, the linear stochastic differential equation which characterizes the hydrological contributions is given by, 

\begin{equation}
\label{ede}
dH_{t}=\alpha (\mu(t)-H_{t})d_{t}+\sigma H_{t}^{\gamma}dB_{t}
\end{equation}

With the initial condition $H_{0}=h$, where $\alpha>0$  is the rate of reversion, $\sigma>0$ is the parameter associated with volatility, $\gamma=0$ are constants that determine the sensitivity of the variance to the level of $H_{t}$, $\left \{B_{t}\right \}_{t\geq 0}$ is a One-dimensional Standard Brownian Motion defined in a probability space $(\Omega,\mathcal{F},\mathbb{P})$, and $\mu(t)$ that is the mean reversion level is defined as a Fourier series of the form,

\begin{equation}
\label{mu}
\mu(t)=\sum_{k=0}^{n}a_{k}\cos(2\pi tk+\phi_k) \hspace{0.4cm} \text{with} \hspace{0.15cm} n=0,1,2,\cdot\cdot\cdot
\end{equation}

Where $a_{k}$ is the amplitude parameter and $\phi_{k}$ the phase parameter. 

In order to estimate the parameters of equation (\ref{ede}) and (\ref{mu}), the methodology proposed by \cite{Monsalve2017} is used. Thus, $\mu(t)$ can be expressed in terms of the expected value of $H_{t}$, then (\ref{ede}) is defined as,

\begin{equation}
\label{eseuler}
dH_{t}=\alpha \left(m(t)+\frac{\stackrel{.}{m}(t)}{\alpha}-H_{t}\right)d_{t}+\sigma H_{t}^{\gamma}dB_{t}
\end{equation}

With $m(t)=E[H_{t}]$ and $\stackrel{.}{m}(t)=\frac{dm(t)}{dt}$. Then using the Euler-Maruyama scheme in (\ref{eseuler}) and defining a new variable $Y_{t}$,

\begin{equation}
Y_{t}=\frac{H_{t}-H_{t-1}-[\alpha(m_{t-1}-H_{t-1})+\stackrel{.}{m}(t)]\Delta}{H_{t-1}^{\gamma}}=\epsilon_{t}  \hspace{0.15cm} ; \hspace{0.4cm} \epsilon_{t} \sim N(0,\sigma^{2}\Delta) 
\end{equation}

Where their maximum likelihood function is given by,

\begin{equation*}
L(\theta|\left\{Y_{t}\right\})=\left(\frac{1}{2\pi\sigma^{2}\Delta}\right)^{T/2}\cdot\exp\left[\frac{-1}{2\sigma^{2}\Delta}\sum_{i=1}^{T}\left(\frac{H_{i}-H_{i-1}-[\alpha(m_{i-1}-H_{i-1})+\stackrel{.}{m}(i-1)]\Delta}{H_{i-1}^{\gamma}}\right)^{2}\right]
\end{equation*}

Solving the problem of maximization of the previous function we obtain the estimates of $\alpha$ and $\sigma$ such that,

\begin{equation*}
\begin{split}
\hat{\alpha}&=\frac{\sum_{i=1}^{T}\left((H_{i}-H_{i-1}-\stackrel{.}{m}(i-1)\Delta)(m_{i-1}-H_{i-1})/H_{i-1}^{2\gamma}\right)}{\sum_{i=1}^{T}\left[(m_{i-1}-H_{i-1})/H_{i-1}^{\gamma})\right]^{2}\Delta}\\
\\
\hat{\sigma}&=\sqrt{\frac{1}{T\Delta}\sum_{i=1}^{T}\left(\frac{H_{i}-H_{i-1}-[\hat{\alpha}(m_{i-1}-H_{i-1})+\stackrel{.}{m}(i-1)]\Delta}{H_{i-1}^{\gamma}}\right)^{2}}
\end{split}
\end{equation*}

Once we have $\hat{\alpha}$, $\hat{\sigma}$ and $\hat{\mu}(t)$, we obtain a better estimate of $\mu(t)$ through the Fourier analysis, such that,

\begin{equation}
\label{representation}
\hat{\hat{\mu}}_{n}=\sum_{k=o}^{L}a_{k}\cos\left[\frac{2\pi kn}{N}+\phi_{k}\right]
\end{equation}

Where $L=\frac{N}{2}$ for $N$ even, $L=\frac{N-1}{2}$ for $N$ and,

\begin{equation*}
\begin{split}
a_{k}&=\left|\hat{M}_{k}\right|=\sqrt{R(\hat{M}_{k})^{2}+I(\hat{M}_{k})^{2}}\\
\phi_{k}&=\arg(\hat{M}_{k})=tan^{-1}\left(\frac{I(\hat{M}_{k})}{R(\hat{M}_{k})}\right)
\end{split}
\end{equation*} 

With $\hat{M}$ as a vector of complex numbers obtained through the Discrete Fourier Transform (DFT) of the signal $\hat{\mu}(t)$, $R(\hat{M})$ the real part of the rectangular representation of $\hat{M}$ and $I(\hat{M})$ the corresponding imaginary part. Thus, we proceed to re-estimate $\alpha$ and $\sigma$ with a mechanism similar to that used to find $\hat{\alpha}$ and $\hat{\sigma}$, such that,

\begin{equation*}
\begin{split}
\hat{\hat{\alpha}}&=\frac{\sum_{i=1}^{T}\left((H_{i}-H_{i-1})(\hat{\hat{\mu}}_{i-1}-H_{i-1})/H_{i-1}^{2\gamma}\right)}{\sum_{i=1}^{T}\left[(\hat{\hat{\mu}}_{i-1}-H_{i-1})/H_{i-1}^{\gamma})\right]^{2}\Delta}\\
\\
\hat{\hat{\sigma}}&=\sqrt{\frac{1}{T\Delta}\sum_{i=1}^{T}\left(\frac{H_{i}-H_{i-1}-\hat{\hat{\alpha}}(\hat{\hat{\mu}}_{i-1}-H_{i-1})\Delta}{H_{i-1}^{\gamma}}\right)^{2}}
\end{split}
\end{equation*}

\subsection{Estimation and Forecast Results}

For each one of the fundamental periods of the hydrological contribution in Colombia an estimation process is carried out as defined in the previous section, achieving a correct estimation of both the external parameters $\Theta=[\alpha,\sigma]$ and the internal parameters $\Phi=[a_{k},\phi_{k}]$ that allow to characterize the dynamics of this variable in the selected periods. Due to the periodic dynamics of the sample, a period will be selected to later execute a forecast process one period forward and contrast it with real data.

The first step in the estimation process is to find the estimates of $m(t)$ and $\stackrel{.}{m}(t)$ from the discrete data for the hydrological contributions. Thus, $\hat{m}(t)$ which can be obtained through different filters and smoothing techniques, is approximated by the Hodrick-Prescott methodology where the smoothing parameter is $\lambda=40000$, whereas $\hat{\stackrel{.}{m}}(t)$ is obtained through a numerical three-point derivation rule. The smoothing parameter is chosen so that the short-term information is correctly captured while maintaining the trend adjustment. Once we have $\hat{m}(t)$ and $\hat{\stackrel{.}{m}}(t)$ we construct a realization of $Y_{t}$ that allows us to find the first estimate of $\alpha$, $\sigma$ and $\mu(t)$, and then from the Fourier analysis to obtain $\hat{\hat{\alpha}}$, $\hat{\hat{\sigma}}$ and $\hat{\hat{\mu}}(t)$.

In the second estimation phase it is necessary to consider an adequate sinusoidal sum, therefore, following methodology proposed by \cite{Monsalve2017}, we choose a number of cosines in the Fourier series that capture sufficient information. Thus, Table \ref{tab:sumcos} shows the RMS of $\hat{\hat{\mu}}(t)$ calculated with L sinusoidal sums in the DFT, with respect to $\hat{\hat{\mu}}(t)$ calculated with L-1 sinusoidal sums in the DFT, the results show that as the number of cosines increases, the RMS tends to decrease. For the period between 2004 and 2007 a total of 26 cosines are sufficient, 24 for the period 2007-2010, 21 for the period 2010-2013 and 20 for the period 2013-2016.

\begin{table}[h]
	\centering
	\caption{\textbf{RMS for L Sinusoidal Sums}}
	\begin{tabular}{ccccccccc}
		\toprule
		\toprule
		\textbf{L-sum} &       & \textbf{2004-2007} &       & \textbf{2007-2010} &       & \textbf{2010-2013} &       & \textbf{2013-2016} \\
		\cmidrule{1-1}\cmidrule{3-3}\cmidrule{5-5}\cmidrule{7-7}\cmidrule{9-9}
		1     &       & 0.038972111 &       & 0.041191016 &       & 0.039792972 &       & 0.041321361 \\
		2     &       & 0.022651407 &       & 0.029692319 &       & 0.035853134 &       & 0.010057239 \\
		3     &       & 0.005210850 &       & 0.024119046 &       & 0.018588931 &       & 0.007456015 \\
		4     &       & 0.004791647 &       & 0.018258372 &       & 0.014054704 &       & 0.003561170 \\
		5     &       & 0.004053573 &       & 0.007316677 &       & 0.008052817 &       & 0.003360246 \\
		$\cdot$     &       & $\cdot$     &       & $\cdot$     &       & $\cdot$     &       & $\cdot$ \\
		15    &       & 0.000045839 &       & 0.000090209 &       & 0.000092941 &       & 0.000262686 \\
		16    &       & 0.000026358 &       & 0.000056475 &       & 0.000072419 &       & 0.000256432 \\
		17    &       & 0.000025099 &       & 0.000052266 &       & 0.000039861 &       & 0.000191860 \\
		18    &       & 0.000022480 &       & 0.000041556 &       & 0.000024837 &       & 0.000160290 \\
		19    &       & 0.000018664 &       & 0.000024320 &       & 0.000018994 &       & 0.000144984 \\
		\textbf{20} &       & 0.000017204 &       & 0.000023895 &       & 0.000018010 &       & \textbf{0.000106069} \\
		\textbf{21} &       & 0.000016727 &       & 0.000021549 &       & \textbf{0.000017071} &       & 0.000071689 \\
		22    &       & 0.000015768 &       & 0.000012234 &       & 0.000006973 &       & 0.000066206 \\
		23    &       & 0.000014560 &       & 0.000010402 &       & 0.000006144 &       & 0.000058468 \\
		\textbf{24} &       & 0.000013902 &       & \textbf{0.000010054} &       & 0.000005167 &       & 0.000057682 \\
		25    &       & 0.000012391 &       & 0.000009897 &       & 0.000004493 &       & 0.000052665 \\
		\textbf{26} &       & \textbf{0.000011072} &       & 0.000009710 &       & 0.000003686 &       & 0.000047684 \\
		27    &       & 0.000009292 &       & 0.000009521 &       & 0.000003607 &       & 0.000047100 \\
		28    &       & 0.000008372 &       & 0.000009348 &       & 0.000003416 &       & 0.000046930 \\
		29    &       & 0.000007855 &       & 0.000008850 &       & 0.000003323 &       & 0.000045595 \\
		30    &       & 0.000007559 &       & 0.000008713 &       & 0.000002763 &       & 0.000044605 \\
		\bottomrule
		\bottomrule
		\multicolumn{9}{p{14.5cm}}{\footnotesize{The RMS of $\hat{\hat{\mu}}(t)$ with L sinusoidal sums, with respect to $\hat{\hat{\mu}}(t)$ with L-1 sinusoidal sums is calculated as $\frac{\sum (\hat{\hat{\mu}}(t)_{L}-\hat{\hat{\mu}}(t)_{L-1})^{2}}{n}$. Calculations are made with $\gamma=0$ and $\Delta t=\frac{1}{365}$ for the hydrological contributions in Colombia during the period 2004 to 2016.}}
	\end{tabular}
	\label{tab:sumcos}
\end{table}

In this sense, Table \ref{tab:paramex} shows the results of the estimation of the external parameters $\alpha$ and $\sigma$ in its two phases, estimation and re-estimation, this calculations were done with the proposed methodology for a $\Delta t=\frac{1}{365}$ since the frequency of the data is daily. The reversion rate $\alpha$ presents values between 90 and 125 for the four periods, being higher in the period between 2004 and 2007, and lower between 2013 and 2016, for this case the re-estimation phase gives lower values. For its part, the parameter of volatility $\sigma$ presents values near 3 for the four periods, where the fourth and third period have the highest values respectively.

\begin{table}[htbp]
	\centering
	\caption{\textbf{Estimation External Parameters}}
	\resizebox{\textwidth}{!}{
		\begin{tabular}{ccccccccccc}
			\toprule
			\toprule
			\textbf{Parameter} &       & \textbf{Phase} &       & \textbf{2004-2007} &       & \textbf{2007-2010} &       & \textbf{2010-2013} &       & \textbf{2013-2016} \\
			\cmidrule{1-1}\cmidrule{3-3}\cmidrule{5-5}\cmidrule{7-7}\cmidrule{9-9}\cmidrule{11-11} 
			\multirow{2}[0]{*}{$\alpha$} &       & \textbf{Estimation} &       & 124.0413 &       & 113.3772 &       & 114.7861 &       & 96.3809 \\
			&       & \textbf{Re-estimation} &       & 121.7219 &       & 112.1324 &       & 112.3681 &       & 91.8261 \\
			&       &       &       &       &       &       &       &       &       &  \\
			\multirow{2}[1]{*}{$\sigma$} &       & \textbf{Estimation} &       & 3.0081 &       & 2.9808 &       & 3.1593 &       & 3.2034 \\
			&       & \textbf{Re-estimation} &       & 3.0129 &       & 2.9835 &       & 3.1655 &       & 3.2161 \\
			\bottomrule
			\bottomrule
			\multicolumn{11}{p{17.5cm}}{\footnotesize{External parameter estimation for (\ref{ede}) that defines the hydrological contributions in Colombia for the 4 fundamental periods. The re-estimation of $\mu(t)$ for the period 2004-2007 was done with a sinusoidal sum of 26 cosines, the period 2007-2010 with 24 cosines, the period 2010-2013 with 21 cosines and the period 2013-2016 with 20 cosines. Calculations are made with $\gamma=0$ and $\Delta t=\frac{1}{365}$.}}
	\end{tabular}}
	\label{tab:paramex}
\end{table}

The Fourier analysis is used for the estimation of $\Phi$ for the four fundamental periods. Thus, Table \ref{tab:paramint} shows the results of the indicator $k$, the amplitude parameter $a_{k}$ and the phase angle $\phi_{k}$, where the number of cosines in the sinusoidal sum is chosen according to the RMS criterion defined above in Table \ref{tab:sumcos}.

\begin{table}
	\centering
	\caption{\textbf{Estimation Internal Parameters}}
	\resizebox{\textwidth}{!}{
		\begin{tabular}{ccccccccccccccccccccccc}
			\toprule
			\toprule
			\multicolumn{5}{c}{\textbf{2004-2007}} &       & \multicolumn{5}{c}{\textbf{2007-2010}} &       & \multicolumn{5}{c}{\textbf{2010-2013}} &       & \multicolumn{5}{c}{\textbf{2013-2016}} \\
			\cmidrule{1-5}\cmidrule{7-11}\cmidrule{13-17}\cmidrule{19-23}    $k$     &       & $a_{k}$    &       & $\phi_{k}$  &       & $k$     &       & $a_{k}$    &       & $\phi_{k}$  &       & $k$     &       & $a_k$    &       & $\phi_{k}$  &       & $k$     &       & $a_{k}$    &       & $\phi_{k}$ \\
			\cmidrule{1-1}\cmidrule{3-3}\cmidrule{5-5}\cmidrule{7-7}\cmidrule{9-9}\cmidrule{11-11}\cmidrule{13-13}\cmidrule{15-15}\cmidrule{17-17}\cmidrule{19-19}\cmidrule{21-21}\cmidrule{23-23}          &       &       &       &       &       &       &       &       &       &       &       &       &       &       &       &       &       &       &       &       &       &  \\
			0     &       & 7.3345 &       & 0.0000 &       & 0     &       & 7.3855 &       & 0.0000 &       & 0     &       & 7.4070 &       & 0.0000 &       & 0     &       & 7.3213 &       & 0.0000 \\
			1     &       & 0.0900 &       & 2.2278 &       & 1     &       & 0.2437 &       & -2.8697 &       & 1     &       & 0.2678 &       & -2.9432 &       & 1     &       & 0.0820 &       & 2.8428 \\
			2     &       & 0.1021 &       & -2.7362 &       & 2     &       & 0.1911 &       & -2.2169 &       & 2     &       & 0.1677 &       & -2.8333 &       & 2     &       & 0.1221 &       & -3.0126 \\
			3     &       & 0.2128 &       & -3.0486 &       & 3     &       & 0.2870 &       & -2.7679 &       & 3     &       & 0.1928 &       & -2.9136 &       & 3     &       & 0.2875 &       & -2.7968 \\
			4     &       & 0.0393 &       & 1.9609 &       & 4     &       & 0.0788 &       & -2.6436 &       & 4     &       & 0.1174 &       & 2.8603 &       & 4     &       & 0.0446 &       & -1.6842 \\
			5     &       & 0.0979 &       & 1.9548 &       & 5     &       & 0.1210 &       & -2.4869 &       & 5     &       & 0.0273 &       & -1.0005 &       & 5     &       & 0.0314 &       & -1.8832 \\
			6     &       & 0.2792 &       & 2.6972 &       & 6     &       & 0.2196 &       & 2.8195 &       & 6     &       & 0.2821 &       & 2.8090 &       & 6     &       & 0.1418 &       & 2.8618 \\
			7     &       & 0.0257 &       & 1.8817 &       & 7     &       & 0.0624 &       & 2.9606 &       & 7     &       & 0.0283 &       & 0.2549 &       & 7     &       & 0.0426 &       & -1.7876 \\
			8     &       & 0.0429 &       & 2.3972 &       & 8     &       & 0.0830 &       & 2.6283 &       & 8     &       & 0.0467 &       & 0.0250 &       & 8     &       & 0.0530 &       & -2.8083 \\
			9     &       & 0.0342 &       & -3.0791 &       & 9     &       & 0.0777 &       & -2.0236 &       & 9     &       & 0.1269 &       & 3.1008 &       & 9     &       & 0.0844 &       & -2.0346 \\
			10    &       & 0.0146 &       & -2.7878 &       & 10    &       & 0.0339 &       & -2.1917 &       & 10    &       & 0.0242 &       & -2.5292 &       & 10    &       & 0.0179 &       & -1.0086 \\
			11    &       & 0.0264 &       & 1.6259 &       & 12    &       & 0.0134 &       & -1.6143 &       & 11    &       & 0.0315 &       & -0.2847 &       & 11    &       & 0.0507 &       & -0.5155 \\
			12    &       & 0.0235 &       & -0.0512 &       & 13    &       & 0.0102 &       & 3.0137 &       & 12    &       & 0.0137 &       & 0.5478 &       & 12    &       & 0.0655 &       & -1.2197 \\
			13    &       & 0.0344 &       & -2.4298 &       & 15    &       & 0.0106 &       & 2.5409 &       & 13    &       & 0.0136 &       & 2.6598 &       & 13    &       & 0.0233 &       & -1.8635 \\
			15    &       & 0.0118 &       & 1.7463 &       & 16    &       & 0.0173 &       & 2.5542 &       & 14    &       & 0.0139 &       & 2.0515 &       & 14    &       & 0.0229 &       & -1.2241 \\
			16    &       & 0.0056 &       & -3.0732 &       & 17    &       & 0.0181 &       & -0.9897 &       & 15    &       & 0.0140 &       & -0.4591 &       & 15    &       & 0.0196 &       & -1.6327 \\
			17    &       & 0.0061 &       & 1.9510 &       & 18    &       & 0.0147 &       & -2.0683 &       & 16    &       & 0.0120 &       & 3.0143 &       & 16    &       & 0.0275 &       & -2.8214 \\
			18    &       & 0.0053 &       & 2.2773 &       & 19    &       & 0.0136 &       & -2.5095 &       & 17    &       & 0.0089 &       & -1.9027 &       & 17    &       & 0.0249 &       & -1.6386 \\
			19    &       & 0.0071 &       & 3.0507 &       & 20    &       & 0.0070 &       & -1.3500 &       & 19    &       & 0.0070 &       & 2.7781 &       & 18    &       & 0.0170 &       & -1.5998 \\
			21    &       & 0.0059 &       & 2.7554 &       & 22    &       & 0.0091 &       & -1.9034 &       & 22    &       & 0.0062 &       & 2.6418 &       & 20    &       & 0.0226 &       & -1.7585 \\
			22    &       & 0.0096 &       & 2.6346 &       & 24    &       & 0.0069 &       & -1.8858 &       & 24    &       & 0.0060 &       & 2.6528 &       & -     &       & -     &       & - \\
			23    &       & 0.0058 &       & 2.6148 &       & 26    &       & 0.0066 &       & -2.1194 &       & -     &       & -     &       & -     &       & -     &       & -     &       & - \\
			24    &       & 0.0067 &       & 1.9361 &       & 27    &       & 0.0046 &       & -1.7297 &       & -     &       & -     &       & -     &       & -     &       & -     &       & - \\
			25    &       & 0.0073 &       & 2.4051 &       & 30    &       & 0.0049 &       & -1.8031 &       & -     &       & -     &       & -     &       & -     &       & -     &       & - \\
			27    &       & 0.0050 &       & 1.9912 &       & -     &       & -     &       & -     &       & -     &       & -     &       & -     &       & -     &       & -     &       & - \\
			28    &       & 0.0054 &       & 2.2135 &       & -     &       & -     &       & -     &       & -     &       & -     &       & -     &       & -     &       & -     &       & - \\
			\bottomrule
			\bottomrule
			\multicolumn{23}{p{20.5cm}}{\footnotesize{Internal parameter estimation for (\ref{representation}) that defines the mean reversion level of the hydrological contributions in Colombia for the 4 fundamental periods. The re-estimation of $\mu(t)$ for the period 2004-2007 was done with a sinusoidal sum of 26 cosines, the period 2007-2010 with 24 cosines, the period 2010-2013 with 21 cosines and the period 2013-2016 with 20 cosines. Calculations are made with $\gamma=0$ and $\Delta t=\frac{1}{365}$.}}
	\end{tabular}}
	\label{tab:paramint}
\end{table}

\begin{figure}[h]
	\centering
	\includegraphics[width=\textwidth]{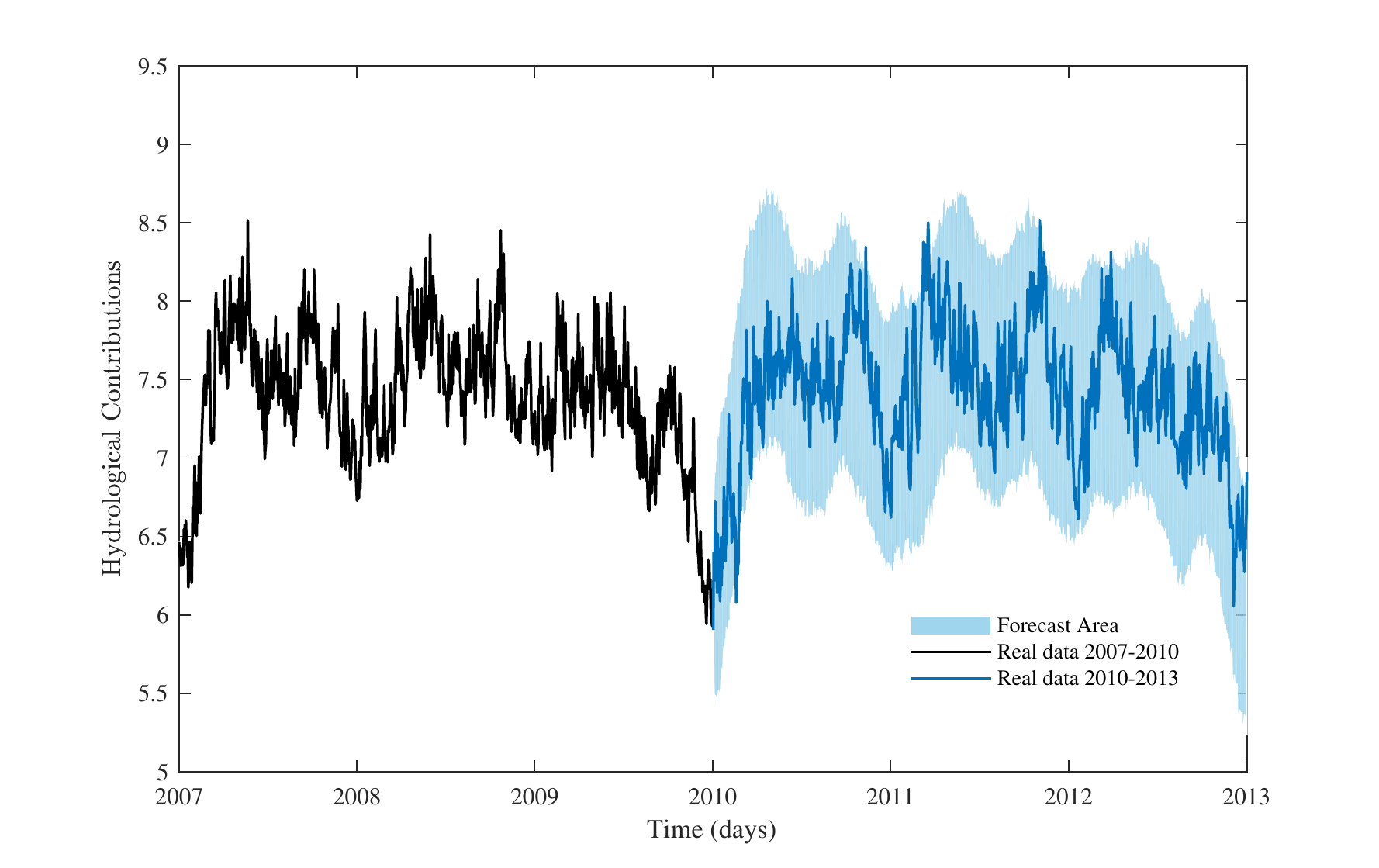}
	\caption{\textbf{Forecast Area with Minimums and Maximums}}
	\medskip
	\begin{minipage}{0.8\textwidth} 
		{\footnotesize{We select the period 2007-2010 then simulate 10000 paths a posterior period with the estimated parameters, of these paths for each point in time we take the minimum and the maximum and construct lower and upper limits.\par}}
	\end{minipage}
	\label{fig:6}
\end{figure}

For the forecasting process, the period from 2007 to 2010 is chosen, given the similarity between the rate of reversion re-estimated ($\hat{\hat{\alpha}}$) for this period and the next. Thus, the parameters estimated from this period are taken to simulate paths in a posterior period, which is equivalent to a three years of forecast for hydrological contributions in Colombia. As first step we estimate 10000 paths from the internal and external parameters found, the calculation of these paths was done through a technique of reduction of variance, such that 5000 trajectories were simulated with positive random values while the remaining 5000 trajectories were simulated with negative random values. Then for each point in time we choose the minimum and maximum of simulated paths, construct lower and upper limits, and adhere the real dynamics of hydrological contributions for the period 2010-2013, as shown in Figure \ref{fig:6}. The results indicate that the proposed forecast band is adjusted efficiently to the real data, since most observations are contained in such bands, while the dispersion of the bands with respect to the actual data is not extensive.

To guarantee efficiency in the forecasting process, we proceeded to form bands made from the best estimate of $\mu$, ($\hat{\hat{\mu}}(t)$), adhering or withdrawing different levels of historical standard deviation of the process, beginning at 0.5 and ending at 2.6. Thus, the upper bands were calculated as ($\hat{\hat{\mu}}(t)+i\sigma_{H_{t}}$) with $i=0.5,\cdots,2.6$ while the lower bands were calculated as ($\hat{\hat{\mu}}(t)-i\sigma_{H_{t}}$) with $i=0.5,\cdots,2.6$. Once we have the bands at different levels of standard deviation, we take the observations of the 10000 trajectories defined with the information of the period 2007-2010 and check point by point how many observations are contained within the bands and how many are outside. We then calculate the probability of each trajectory to be within the bands and obtain the average of such probabilities at different levels of standard deviation, as it is denoted in column 2 of Table \ref{tab:probab}, column 3 for its part, corresponds to the probability that the real data for the period 2010-2013 are contained in the previously defined bands, while column 4 is only the difference between column 3 and column 2.

\begin{table}[h]
	\centering
	\caption{\textbf{Probability of Occurrence}}
	\begin{tabular}{ccccccc}
		\toprule
		\toprule
		\textbf{Standard D.} &       & \textbf{Forecast} &       & \textbf{2010-2013} &       & \textbf{Difference} \\
		\cmidrule{1-1}\cmidrule{3-3}\cmidrule{5-5}\cmidrule{7-7}
		0.5   &       & 67.80\% &       & 47.63\% &       & -20.17\% \\
		0.6   &       & 76.54\% &       & 55.20\% &       & -21.34\% \\
		0.7   &       & 83.44\% &       & 62.86\% &       & -20.57\% \\
		0.8   &       & 88.69\% &       & 70.26\% &       & -18.43\% \\
		0.9   &       & 92.53\% &       & 76.19\% &       & -16.34\% \\
		1.0   &       & 95.24\% &       & 81.93\% &       & -13.30\% \\
		1.1   &       & 97.07\% &       & 85.22\% &       & -11.85\% \\
		1.2   &       & 98.25\% &       & 88.87\% &       & -9.38\% \\
		1.3   &       & 98.99\% &       & 90.97\% &       & -8.03\% \\
		1.4   &       & 99.44\% &       & 93.25\% &       & -6.19\% \\
		1.5   &       & 99.70\% &       & 94.89\% &       & -4.81\% \\
		1.6   &       & 99.85\% &       & 95.62\% &       & -4.23\% \\
		1.7   &       & 99.93\% &       & 96.72\% &       & -3.21\% \\
		1.8   &       & 99.96\% &       & 97.72\% &       & -2.25\% \\
		1.9   &       & 99.98\% &       & 97.99\% &       & -1.99\% \\
		2.0   &       & 99.99\% &       & 98.63\% &       & -1.36\% \\
		2.1   &       & 100.00\% &       & 98.72\% &       & -1.27\% \\
		2.2   &       & 100.00\% &       & 99.09\% &       & -0.91\% \\
		2.3   &       & 100.00\% &       & 99.54\% &       & -0.46\% \\
		2.4   &       & 100.00\% &       & 99.64\% &       & -0.36\% \\
		2.5   &       & 100.00\% &       & 99.91\% &       & -0.09\% \\
		2.6   &       & 100.00\% &       & 100.00\% &       & 0.00\% \\	    
		\bottomrule
		\bottomrule
		\multicolumn{7}{p{11cm}}{\footnotesize{The Forecast column is calculated as the average of the probabilities that the 10000 paths simulated from the information for the period 2010-2013 are contained in the lower and upper bands calculated as ($\hat{\hat{\mu}}(t)\pm i\sigma_{H_{t}}$) with $i=0.5,\cdots,2.6$. The 2010-2013 column is calculated as the probability that this period contained in the previous bands. The Difference column is calculated as the difference between column 3 and column 2.}}
	\end{tabular}
	\label{tab:probab}
\end{table}

The results indicate, as expected, that with greater standard deviations, the probability that the path of the real data and the simulated paths are located in the forecast bands is greater. For the 10000 simulated paths with information from the 2007-2010 period, all data are contained in the confidence band made with 2.1 standard deviations. For its part, the real data for the period 2010-2013 are contained entirely in the confidence band made with 2.6 standard deviations. Note that from 2.1 standard deviations the difference between the probability that the real data and the simulated data are contained in the confidence band is small (-0.91$\%$), therefore it can be affirmed that the forecast method is efficient even when standard deviation levels below 2.1 are considered. 

\begin{figure}[h]
	\centering
	\includegraphics[width=\textwidth]{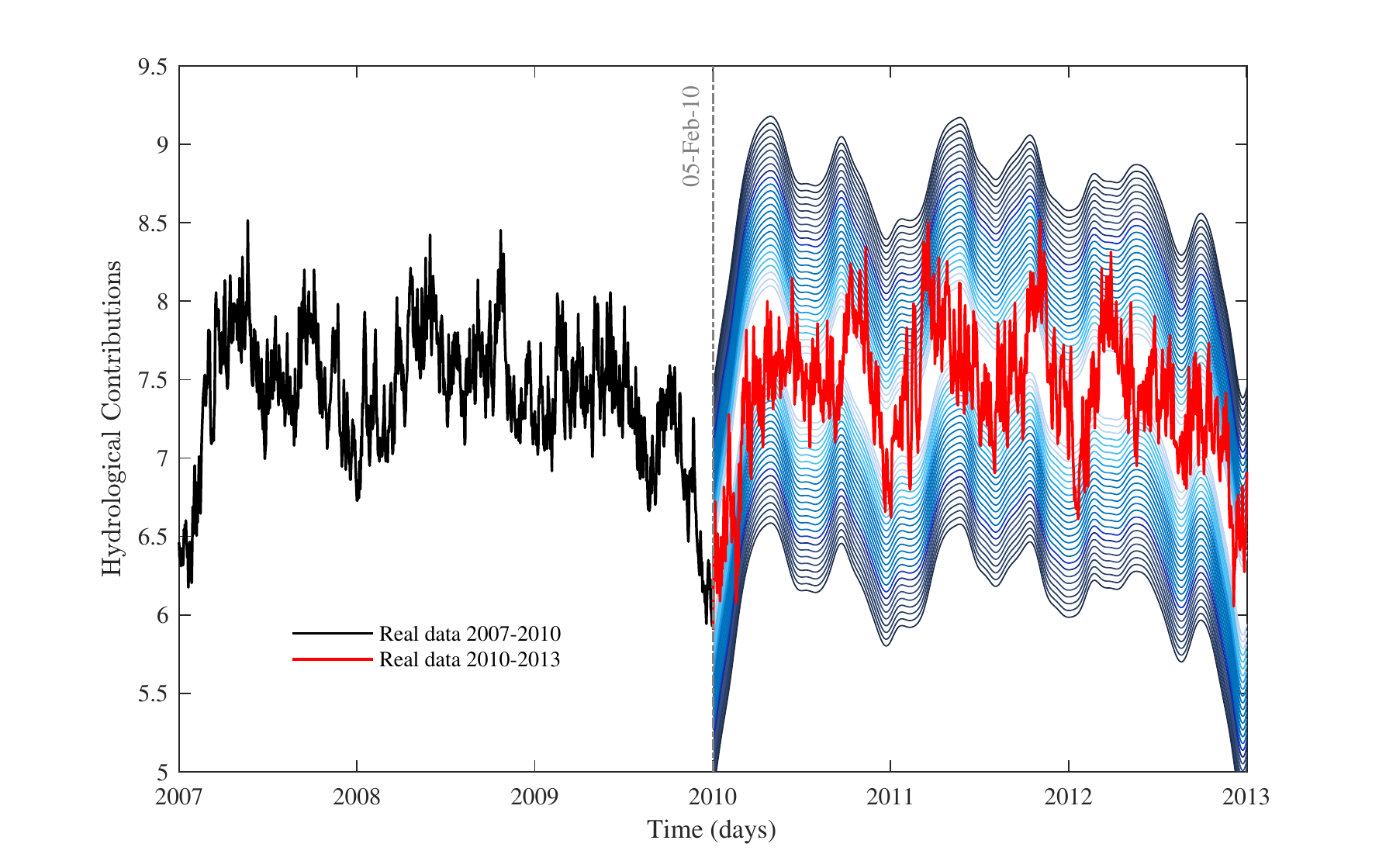}
	\caption{\textbf{Bands of Forecast to Different Levels of Standard Deviation}}
	\medskip
	\begin{minipage}{0.8\textwidth} 
		{\footnotesize{We select the period 2007-2010 and construct forecast bands based on the information contained in this period. The bands are made at different levels of standard deviation and are calculated as $\hat{\hat{\mu}}(t)\pm i\sigma_{H_{t}}$ with $i=0.5,\cdots,3$.\par}}
	\end{minipage}
	\label{fig:7}
\end{figure}

Graphically, these results are shown in Figure \ref{fig:7}. In this figure, forecast bands are observed at different levels of historical standard deviation of the process, calculated as $\hat{\hat{\mu}}(t)\pm i\sigma_{H_{t}}$ with $i=0.5,\cdots,3$, and contrasted with the real data. Thus, at higher standard deviations, the forecast bands capture most of the observations of the real data, particularly from 2 standard deviations.

\section{Conclusions and Comments}\label{conclusions}

The generation of electric energy in Colombia is mainly based on hydraulic generation due to the hydrological potential of the country. Most of the hydroelectric plants are located in the Andean, Caribbean and Pacific regions, which make use of surface water flows from such regions. However, this hydrological potential is influenced by atmospheric conditions, inter-annual events such as ENSO, environmental change by anthropogenic action, among others, causing the water level in the reservoirs to vary substantially and, therefore, the prices of the electric energy. Under this scenario, modeling and forecasting the variables that affect the hydric resource, is a work of great importance for agents involved in the colombian electricity sector, particularly for generators, since it allows them to efficiently manage the generation process of electric energy.

One of the variables of hydrological character with greater relevance corresponds to the water discharge of the rivers that contribute water to the reservoirs of the SIN. Such variable measured in $m^{3}/s$ is a proxy of the hydric supply for each reservoir and for the system in aggregate terms. These water discharge or hydrological contributions in their first logarithmic transformation for the period between 2004 and 2016 have exhibited periodic dynamics in daily frequency with a fundamental period that is repeated every 3 years, with sub-periods that are repeated each year.

Using this periodicity we employ the maximum likelihood estimation and the Fourier analysis through the Discrete Fourier Transform (DFT), to estimate the parameters of one-factor mean reversion stochastic process where the functional trend follows to periodic behavior, to subsequently make a forecast of a specific period (2010-2013) with the information of the period immediately preceding. In this sense, the estimation method proposed is efficient and useful to characterize the dynamics of hydrological contributions in the colombian SIN. The forecast for its part, presents results were close to the real data, particularly, the forecast bands constructed with the periodic functional trend and with approximately 2 standard deviations contain the majority of the observations of the real data.

The proposed method allows the agents of the Mercado de Energía Mayorista (MEM) to analyze the dynamics of the hydric resource on a daily frequency, to take the best decisions in the management of their assets. In addition, the forecast in question provides a time window of approximately 3 years, giving greater advantages in contrast to traditional forecasting methods.

\section*{Acknowledgments}

To the company XM for giving us all the available information regarding the water discharge of the rivers belonging to the SIN. Empresas Públicas de Medellín (EPM), EMGESA, Empresa de Energía del Pacífico (EPSA), Empresa URRÁ and ISAGEN provided information on the water discharge for the rivers that supply water to the reservoirs for which they are owners or operators.

\bibliography{library}

\end{document}